\documentclass[10pt,showkeys,showpacs,a4paper,aps,pra,amssymb]{revtex4-1} 
\usepackage{amsmath}
\usepackage{amsfonts}
\usepackage{textcomp}
\usepackage{graphicx}
\usepackage{algorithm}
\usepackage{algpseudocode}
\usepackage[english]{babel} 
\usepackage[latin1]{inputenc}
\usepackage[colorlinks=true, allcolors=blue]{hyperref} 
\usepackage[all]{hypcap}   
\usepackage{bm}
\usepackage{natbib}  
\usepackage[usenames, dvipsnames]{color} 
\let\originaleqref=\eqref
\renewcommand{\eqref}{equation~\originaleqref}
\scrollmode
\begin{document}

\title{Study of plasmonic slot waveguides with a nonlinear metamaterial core: semi-analytical and numerical methods}
\author{Mahmoud M. R. Elsawy}
\author{Gilles Renversez$^*$}
\affiliation{Aix--Marseille Univ, CNRS, Ecole Centrale Marseille, Institut Fresnel, 13013 Marseille, France}
\email[]{gilles.renversez@univ-amu.fr}

\date{\today}

\begin{abstract}
Two distinct models  are developed to investigate the transverse magnetic stationary solutions propagating in one-dimensional anisotropic nonlinear plasmonic structures made from a nonlinear metamaterial core of Kerr-type embedded between two semi-infinite metal claddings. The first model is semi-analytical, in which we assumed that the anisotropic nonlinearity depends only on the transverse component of the electric field and that the nonlinear refractive index modification is small compared to the linear one. This method allows us to derive analytically the field profiles and the nonlinear dispersion relations in terms of the Jacobi elliptical functions. The second model is fully numerical, it is based on the finite-element method in which all the components of the electric field are considered in the Kerr-type nonlinearity with no presumptions on the nonlinear refractive index change. Our finite-element based model is valid beyond weak nonlinearity regime and generalize the well-known single-component fixed power algorithm that is usually used. Examples of the main cases are investigated including ones with strong spatial nonlinear effects at low powers. Loss issues are reduced through the use of gain medium in the nonlinear metamaterial core.
\end{abstract}

\maketitle
\thispagestyle{plain}
\section{Introduction}
Nonlinear optical properties play a crucial role in all-optical integrated circuits due to the different control functionalities they offer~\cite{Tien:71-integrated-optics,stegeman1985nonlinear-integrated-optics,stegeman1988third-nl-integrated}. Utilizing plasmonics as a part of nonlinear structures may be a promising choice because of the reduced footprint achievable compared with all-dielectric structures, and because of the enhancement of the field intensities, which can be used to boost the nonlinearity~\cite{kauranen_nonlinear_2012,rukhlenko_nonlinear_2011}. Several nonlinear plasmonic waveguides  have already been studied~\cite{Ariyasu85,Agranovich80,Walasik14,ajith2015linear-Metal-NL}. The structures composed of a nonlinear isotropic core of Kerr-type sandwiched between two semi-infinite metal claddings, have received a great attention since their study in 2007~\cite{Feigenbaum07}, due to the strong light confinement obtained and due to its peculiar nonlinear effects~\cite{Davoyan08,rukhlenko2011exact-dispersion,Ferrando13,salguiero14complex-modes-plasmonic-nonlinear-slot-waveguides,Walasik15b}. 
 These nonlinear plasmonic slot waveguides (NPSWs) promise a family of exciting applications such as phase matching in higher harmonic generation processes~\cite{Davoyan09a}, nonlinear plasmonic couplers~\cite{sammut_theoretical_1993} or switching~\cite{Nozhat12}. In order to study light propagating in such structures, different methods have been developed~\cite{Davoyan08,rukhlenko2011dispersion,rukhlenko2011exact-dispersion} to describe its main modes. Recently, a full description of the solutions including the higher-order ones was introduced~\cite{Walasik15a,Walasik15b}. In addition, an improvement of such structures by the inclusion of two linear  dielectric buffer layers between the nonlinear core and the two metal claddings has also been proposed~\cite{Elsawy_OL_16,Elsawyspie15}. Nevertheless, all these studies deal only with the standard nonlinear isotropic core with focusing Kerr-type. It appears that the power needed to observe the symmetry breaking of the fundamental symmetric mode is in the range of GW/m which is still too high. Otherwise, during the last few years, it was demonstrated that the nonlinear effects can be extremely enhanced using epsilon-near-zero (ENZ) materials~\cite{ciattoni2010extreme,campione2013electric-field-enhancement-ENZ,ciattoni2016enhanced,neira2015eliminating,alam2016large-indium-tin-oxide-ENZ}. However, even if recent experimental results  have demonstrated a large enhancement of third harmonic generation in the epsilon-near-zero (ENZ) regime~\cite{Capretti15OL-enhanced-third-harmonic-generation}, the usual modelling approaches are no longer applicable in this high nonlinear regime since the spatial  nonlinearity cannot be treated as a perturbation. We have shown in a recent work that in order to fully take advantage of the ENZ nonlinearity enhancement in nonlinear waveguides, it is important to introduce anisotropy~\cite{Elsawy16-OL-anisotropic-metamaterial-waveguides}. Even if, plasmonic waveguides have already been studied with metamaterial layers either in the cladding~\cite{ishii2014plasmonic-HMM-cladding} or in the core~\cite{avrutsky2007-HMM-core-linear,rukhlenko2012guided-plasmonic-anisotropic}, these last studies focused only on linear guided waves. In~\cite{ciattoni2010extreme}, nonlinear guided waves were studied in anisotropic media, but with an effective nonlinear response identical to the standard isotropic one for transverse magnetic (TM) polarized waves. In our study, we consider a nonlinear metamaterial with an anisotropic effective response for the TM polarized waves. To allow this approach, we generalize two of our previous methods to tackle anisotropic nonlinear cores and not only isotropic ones. For the first method, this implies both a new classification of the possible cases for the effective nonlinearity and more complicated nonlinear dispersion relations compared with the isotropic case. The second method is based on the finite element method and the fixed power algorithm where the nonlinear stationary solutions are computed numerically as a function of the fixed input power. Usually, it is assumed that only the transverse component of the electric field is taken into account in the nonlinear form. Here, this is not the case, since we take into account all the electric field components in the Kerr-type nonlinearity. In order to achieve this result, we introduce and solve two coupled scalar nonlinear equations, on per the continuous tangential component of the electromagnetic field.

 In this study, we present the full derivation of two different methods to study nonlinear TM stationary solutions propagating in one-dimensional plasmonic slot waveguides with an anisotropic metamaterial nonlinear core. The first one is semi-analytical while the second model is fully numerical and is based on the finite-element method (FEM).
 
These models have been briefly introduced in~\cite{Elsawy16-OL-anisotropic-metamaterial-waveguides}, while here we present their detailed derivations.
The article is organized as follows, in section~\ref{sec:problem} the statement of the problem is presented. The semi-analytical method and the numerical FEM are described in section~\ref{sec:derivation_models}. In section~\ref{sec:numerical_results} the validation of our models is given by adapting them to study the isotropic case and the comparison with the results from previously published work is described. In addition, we present several examples to show  the influence of the anisotropic nonlinearity on the nonlinear dispersion curves and on the field profiles. 
\section{Problem formalism}
\label{sec:problem}
\begin{figure}[htbp]
\centering
\includegraphics[width=0.7\columnwidth]{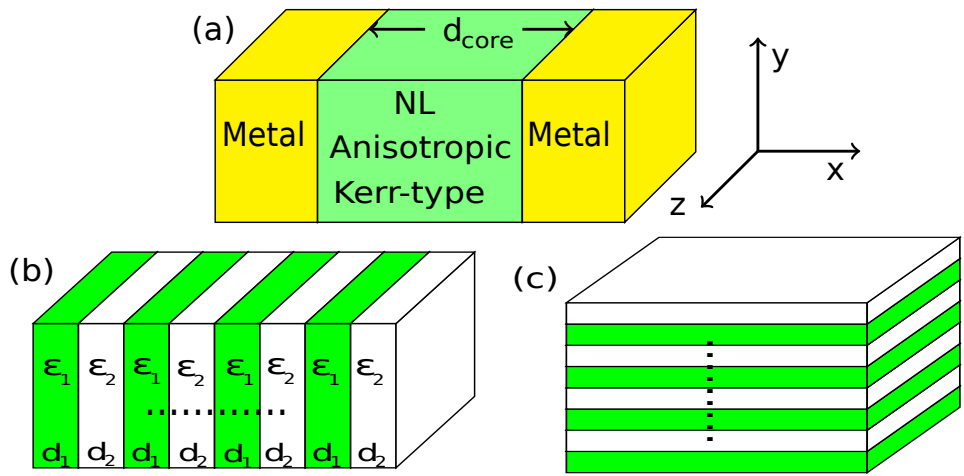}
\caption{(a): Symmetric nonlinear plasmonic slot waveguide geometry with its metamaterial nonlinear core and the two semi-infinite metal claddings. (b) and (c): Two different orientations for the nonlinear metamaterial core formed by periodic layers of two different isotropic media.}
\label{fig:structure}
\end{figure}

We consider a structured metamaterial nonlinear core, formed by bulk layers, embedded between two semi-infinite isotropic metal claddings (see figure~\ref{fig:structure}(a)). The core layers consist of two different isotropic media with two different permittivities $\epsilon_{1}$, $\epsilon_{2}$ and two different thicknesses $d_{1}$, $d_{2}$, respectively as shown in figure~\ref{fig:structure}(b). We propose two different models to study light propagation in such anisotropic nonlinear structures.
The first model is based on the approach presented in~\cite{Walasik15a,Walasik15b} for isotropic nonlinear core, extending it to structures containing an anisotropic nonlinear core. In this method, the nonlinearity is treated in an approximated way such that all the components of the permittivity tensor depend only on the transverse component of the electric field. Another limitation is that the nonlinear refractive index change is small compared to the linear refractive index. Based on the above assumptions, analytical formulas for the field profiles in terms of the Jacobi elliptical functions~\cite{Abramowitz-Handbook-Mathematical-functions,Byrd54-Handbook-elliptic-integrals} are obtained; with the continuity of the tangential electromagnetic field components, they allow us to derive analytical formulas for the nonlinear dispersion relations. This method will be called the extended Jacobi elliptical model (EJEM). Our EJEM, is different from the simple model used in references~\cite{Ariyasu85,Stegeman85}, in which only two components of the  permittivity tensor depend on the transverse component of the electric field, while in our case we consider the dependence of the transverse component of the electric field for all the components of the permittivity tensor. In addition, the methods used in~\cite{Ariyasu85,Stegeman85}, developed to study stationary solutions propagating in structures containing a semi-infinite nonlinear region, whereas in our case we consider a finite-size nonlinear anisotropic core. The second model described in the current study is fully numerical and based on the FEM to solve the stationary TM  problem in nonlinear layered structures. This numerical FEM does not require any of the assumptions used in the semi-analytical EJEM, and all the components of the effective nonlinear permittivity tensor depend on both the transverse and the longitudinal components of the electric field, moreover the nonlinear term does not need to be small, which means that it is valid beyond the weak nonlinearity regime. In order to treat the nonlinearity in the FEM, we generalize the fixed power algorithm presented in~\cite{Drouart08,Walasik14} and we consider a coupled nonlinear eigenvalue problem to take into account all the electric field components in the Kerr-type nonlinearity instead of the single scalar eigenvalue problem solved previously. Furthermore, in our new model, the nonlinearity is treated without any assumptions relative to its amplitude.
It is worth mentioning that the first semi-analytical model (EJEM), provides more insight and understanding into the nature of the of finding stationary solutions in the structure than the second, more numerical model. Nevertheless, the second model treats the nonlinearity in a proper way without any assumptions on the Kerr-nonlinearity amplitude, but the field profiles are computed numerically.

Our models are written for TM light polarization in which the magnetic field has only one component such that $\pmb{\mathcal{H}}=\left[0,\mathcal{H}_{y},0 \right]$ and the electric field has two components $\pmb{\mathcal{E}}=[\mathcal{E}_{x},0,i\mathcal{E}_{z}]$. We consider only monochromatic TM waves propagating along the $z$-direction in one-dimensional symmetric structures which are invariant along $z$ and $y$ directions (see figure~\ref{fig:structure}(a)). In these structures, all the field components evolve proportionally to $exp[i(k_{0} n_{eff} z-wt)]$: $\pmb{\mathcal{E}}(x,z,t)=\pmb{E}(x) exp[i(k_{0} n_{eff} z-wt)]$, $\pmb{\mathcal{H}}(x,z,t)=\pmb{H}(x) exp[i(k_{0} n_{eff} z-wt)],$ 

where $k_{0}=w/c$ represents the wave number in vacuum, $\omega$ denotes the angular frequency of the wave, and c is the speed of light in vacuum. Here $n_{eff}$ and $k_{0} n_{eff}$ denote the effective index and the propagation constant of the wave, respectively. 
The magnetic field induction vector is defined as $\pmb{\mathcal{B}}=\mu_{0} \pmb{\mathcal{H}}$, where the permeability is constant and is denoted by $\mu_{0}$, that of vacuum. 
In our models, we assume that the relative permittivity tensor is complex and diagonal: $\tilde{\bar{\bar{\epsilon}}}(x)=\bar{\bar{\epsilon}}(x)+i\bar{\bar{\epsilon}}^{\prime \prime}(x)$. We consider only the real part of the permittivity $\bar{\bar{\epsilon}}(x) =[\epsilon_{x} ~~ \epsilon_{y} ~~ \epsilon_{z}]$ in the derivation of the nonlinear dispersion relations, such that the displacement vector is defined as $\pmb{\mathcal{D}}=\epsilon_{0} {\bar{\bar{\epsilon}}} \pmb{\mathcal{{E}}}$, with $\epsilon_{0}$ being the vacuum permittivity, the imaginary parts will be used to estimate the losses, in which we use the same method as in~\cite{Davoyan09,Walasik14,Stegeman85}, but extended to the anisotropic case (see section~\ref{sec:numerical_results}).
Depending on the chosen orientation of the compound layers relative to the Cartesian coordinate axes, different  anisotropic permittivity tensors can be build for the core. Due to the $z$-invariance hypothesis required by our modelling, only two types where the $z$-axis belongs to the layers have to be considered. For the first one  where the layers are parallel to the $y$-axis (figure~\ref{fig:structure}(b)), one has for the diagonal terms of ${\bar{\bar{\epsilon}}}$: $[ \varepsilon_{x}  =\varepsilon_{\perp}  \;\,  \varepsilon_{y}= \varepsilon_{//}   \;\, \varepsilon_{z}=\varepsilon_{//}] $. For the second case where the layers are parallel to the $x$-axis, one gets: $[ \varepsilon_{x}  =  \varepsilon_{//} \;\,  \varepsilon_{y}  =  \varepsilon_{\perp}   \;\,   \varepsilon_{z}  =  \varepsilon_{//}]$ (see figure~\ref{fig:structure}(c)). The first one allows us to obtain an anisotropic effective nonlinear response for TM polarized waves, unlike the second case in which the effective nonlinear response coincides with the standard isotropic one. We will focus on the first orientation of the layers as depicted in figure~\ref{fig:structure}(b).
The nonlinearity considered in this study is the usual Kerr-type, where all the components of the relative permittivity tensor depend on the electric field intensity:
\begin{equation}
\label{eq:real_permitt_tensor}
{ {\epsilon_{j}}}={{\epsilon_{jj}}}+{ {\alpha_{jj}}}  | {\pmb{E}}(x) |^{2}~~~\forall j \in \{x,y,z \},
\end{equation}
in which ${ {\epsilon_{jj}}}$ is the j-th component of the real linear permittivity tensor and ${{\alpha_{jj}}}$ is the corresponding nonlinear parameter, $\forall j \in \{x,y,z \}$.  This kind of nonlinearity has already been used in many nonlinear waveguide studies~\cite{Feigenbaum07,Agranovich80,Maradudin88,Walasik15a}, as a first step we do not need to consider the complex Kerr-type nonlinearity used in~\cite{ciattoni2010extreme,rizza2011two-peaked-flat-top}. 

Finally, using the definition of the magnetic field induction $\pmb{\mathcal{B}}$, the displacement vector $\pmb{\mathcal{D}}$ 
we can express Maxwell's equations for the TM polarized waves as:
\begin{subequations}
\label{eq:maxwell_TM}
\begin{gather}
        E_{x}(x)= \frac{ n_{eff} H_{y}(x)}{\epsilon_{0}\epsilon_{x}(x)c}, \label{eq:Ex_full}\\        
        E_{z}(x)= \frac{1}{\epsilon_{0}\epsilon_{z}(x)\omega} \frac{d H_{y}(x)}{d x}, \label{eq:Ez_full} \\        
        k_{0} n_{eff} E_{x}(x)- \frac{d E_{z}(x)}{d x} =\omega\mu_{0} H_{y}(x). \label{eq:Ex_Ez_Hy}
\end{gather}
\end{subequations}
\section{Derivation of the Methods}
\label{sec:derivation_models}
\subsection{Extended Jacobi elliptical model (EJEM)}
\label{subsec:EJEM}
We begin with the derivation of the field profiles and the nonlinear dispersion relation in the frame of the EJEM, in which strong assumptions on the form of the Kerr-type nonlinearity are required to establish it. This method is a generalization and an extension to the Jacobi elliptical model (JEM) developed to study the stationary nonlinear solutions in isotropic plasmonic slot waveguides~\cite{Walasik15a,Walasik15b}, and based on the same assumptions: (i) the nonlinearity depends only on the transverse component of the electric field and (ii) the nonlinear permittivity modifications are small compared to the linear refractive index such that $\forall j \in \{x,y,z \}$:
\begin{equation}
\label{eq EJEM_assumptions}
\begin{aligned}
\epsilon_{j}=\epsilon_{jj}+\alpha_{jj} |E_{x}|^{2}, \\
\alpha_{jj} |E_{x}|^{2} << \epsilon_{jj}.~~~~
\end{aligned}
\end{equation} 
These assumptions are valid only at low power as it was shown in~\cite{Walasik15a,Walasik14,Maradudin88,Stegeman84,Agranovich80}. However, they allow us to derive analytical formulas for the field profiles inside the anisotropic nonlinear core in terms of the Jacobi elliptical functions~\cite{Abramowitz-Handbook-Mathematical-functions,Byrd54-Handbook-elliptic-integrals} which will be used together with the field in the linear metal claddings and the continuity of the tangential components at the core interface to acquire analytical expressions for the nonlinear dispersion relations.

In order to derive the nonlinear wave equation in terms of the magnetic field component $H_{y}$ in the frame of the EJEM assumptions, we use equations~\ref{eq:maxwell_TM} and proceeding as in the isotropic case~\cite{Walasik14}, we find
\begin{equation}\label{eq:nl_wave_eq_FBM_implicit} 
\frac{d^2 H_{y}}{d x^2} - k_{0}^{2}  q^{2}(x)  H_{y}(x) + k_{0}^2 a^{\texttt{\tiny {EJEM}}}_{nl}(x) H_{y}^3(x)=0, 
\end{equation}
where
\begin{equation}
\label{eq:q_sqr_different_layers}
 q^{2}(x)= 
\left\{
\begin{array}{ll}
      q_{core}^{2}= \left( \frac{[\Re e(n_{eff})]^2 \epsilon_{zz} }{\epsilon_{xx}}-\epsilon_{zz} \right)  ~~~~~~$in the core$,	\\
      q_{m}^{2}   =  \left( [\Re e(n_{eff})]^{2}-\epsilon_{m}\right)        ~~~~~~~~~$in the claddings$,      
\end{array} 
\right. 
\end{equation}
in which $\epsilon_{m}<0$ being the real part of the permittivity in the metal claddings. In equation~\eqref{eq:q_sqr_different_layers}, we consider only the real part of the effective index $\Re e(n_{eff})$ as used in~\cite{Walasik15a,Ariyasu85,Stegeman85}. The nonlinear coefficient $a^{\texttt{\tiny{EJEM}}}_{nl} (x)$ is null in the linear metal claddings while in the nonlinear anisotropic core it is given by:
\begin{equation}\label{eq:a_nl_ansio}
a^{\texttt{\tiny{EJEM}}}_{nl}=\frac{-[\Re e(n_{eff})]^2 }{\epsilon_{xx}^{4} c^{2} \epsilon_{0}^2} \left( [\Re e(n_{eff})]^{2} \left( \alpha_{zz}\epsilon_{xx}-\alpha_{xx}\epsilon_{zz}\right)-\alpha_{zz}\epsilon_{xx}^{2} \right). 
\end{equation}
Our model can be reduced to the isotropic state~\cite{Walasik14,Walasik15a,Walasik15b} by setting $\epsilon_{xx}=\epsilon_{zz}=\epsilon_{l,core}$ and $\alpha_{xx}=\alpha_{zz}=\alpha$ in equations~(\ref{eq:q_sqr_different_layers}) and~(\ref{eq:a_nl_ansio}) from which we recover the same expressions for $q^2$ and $a^{\texttt{\tiny{EJEM}}}_{nl}$ as in the JEM (the same isotropic response can be obtained with the orientation shown in figure~\ref{fig:structure}(c)). It is worth mentioning that the nonlinear coefficient given by~\eqref{eq:a_nl_ansio} is different from what has been developed for the anisotropic case presented in~\cite{Ariyasu85,Stegeman85}. In these two references (using to our notations to facilitate the comparison), it was assumed that $\epsilon_{x}$ and $\epsilon_{y}$ depend on the transverse component of the electric field, while $\epsilon_{z}$ is assumed not to depend on this component (it is assumed to be constant). In the present study, we consider the dependence of the transverse component of the electric field for all the components of the permittivity tensor. As a special case, in our recent work~\cite{Elsawy16-OL-anisotropic-metamaterial-waveguides}, we presented the results considering isotropic nonlinear term ($\alpha_{xx}=\alpha_{zz}>0$) with an anisotropic linear one ($\epsilon_{xx} \neq \epsilon_{zz}$) in order to show the influence of the anisotropic linear part of the permittivity on the nonlinear dispersion diagrams and on the field profiles. In~\cite{Elsawy16-OL-anisotropic-metamaterial-waveguides}, we distinguished between two different cases; the elliptical case, in which $\epsilon_{xx}>0$ and $\epsilon_{zz}>0$, and the hyperbolic case, where  $\epsilon_{xx}>0$ and $\epsilon_{zz}<0$. While, in the present study, we present a general and full treatment for anisotropic linear and nonlinear permittivity terms with more general classifications based on the sign of $a^{\texttt{\tiny{EJEM}}}_{nl}$, which depends on both the effective linear and nonlinear permittivity terms (see~\eqref{eq:a_nl_ansio}). 	
In order to solve~\eqref{eq:nl_wave_eq_FBM_implicit}, we use the first integral approach~\cite{Maradudin88,Mihalache87,Walasik14} 
and integrate in each of the structure layers separately. Consequently, we can write the nonlinear wave equation as:
\begin{align}\label{eq:first_integral_nl_wave_eq_FBM_implicit} 
\left( \frac{d H_{y}}{d x} \right)^{2} - k_{0}^{2}  q^{2}(x)  H^{2}_{y}(x) + k_{0}^2 \frac{a^{\texttt{\tiny {EJEM}}}_{nl}(x)}{2} H_{y}^4(x)=C_{0}, 
\end{align}
where $C_{0}$ is a constant of integration. In the semi-infinite metal cladding, $C_{0}$ is null since the magnetic field and its derivative tends to zero as $x \rightarrow \pm \infty$. Moreover, in the linear cladding, the nonlinear parameter $a^{\texttt{EJEM}}_{nl}=0$ and~\eqref{eq:nl_wave_eq_FBM_implicit} reduces to the standard linear wave equation and the magnetic field matches the usual decaying exponential in the metal regions  whose solutions are given by:
\begin{equation}
\label{eq:Hy_linear_cladding}
\begin{aligned}
~~H_{y}=H_{[-{d_{core}}/{2}]} e^{k_{0}q_{m}(x+\frac{d_{core}}{2})} ~~~~~~\text{for}~~~~  -\infty \leq x < -\frac{d_{core}}{2},\\
H_{y}=H_{[{d_{core}}/{2}]} e^{-k_{0}q_{m}(x-\frac{d_{core}}{2})}  ~~~~~\text{for}~~~  \frac{d_{core}}{2} \leq x < +\infty.
\end{aligned}
\end{equation} 
Here, $H_{[-{d_{core}}/{2}]}$ and $H_{[{d_{core}}/{2}]}$ are the values of the magnetic field amplitudes at the left and the right interfaces, respectively.
We are searching for guided wave solutions in the anisotropic waveguides depicted in figure~\ref{fig:structure}, consequently, we will look only for the solutions with positive attenuation coefficient $q_{m}$ in the metal claddings (see \eqref{eq:Hy_linear_cladding}). While, the quantity $q_{core}$  in the core can be either real or imaginary leading to positive or negative values of $q_{core}^2$ (see~\eqref{eq:q_sqr_different_layers}).

In the nonlinear core, $C_{0} \neq 0$  and this constant can be obtained in terms of the magnetic field amplitudes at the core interfaces using the continuity of the longitudinal components at the core interface such that
\begin{align}
\label{eq:C0_cont_conditions}
C_{0}=k_{0}^{2}H_{[-{d_{core}}/{2}]}\left[\left( \frac{\epsilon_{zz}}{\epsilon_{m}} \right)^{2}q_{m}^{2}-q_{core}^2+\frac{a^{\texttt{\tiny {EJEM}}}_{nl}}{2}H^{2}_{[-{d_{core}}/{2}]}\right].
\end{align}
A similar expression can be obtained for $C_{0}$ at the right interface. It is important to mention that the  value of $\epsilon_{z}$ at the left interface is replaced by $\epsilon_{zz}$ in~\eqref{eq:C0_cont_conditions} since we assumed, in the EJEM, that the nonlinear refractive index change is small compared to the linear one. The expression of $C_{0}$ is a generalization to what have been obtained in the isotropic case~\cite{Walasik15a} with different values of $q^{2}_{core}$ and $a^{\texttt{\tiny{EJEM}}}_{nl}$, such that we recover the same expression in the isotropic case by setting $\epsilon_{xx}=\epsilon_{zz}=\epsilon_{l,core}$ and $\alpha_{xx}=\alpha_{zz}=\alpha$ . Due to the anisotropy, the sign of the nonlinear parameter $a^{\texttt{\tiny{EJEM}}}_{nl}$ could be positive or negative depending on the values of $\epsilon_{xx}$, $\epsilon_{zz}$, $\alpha_{xx}$, and $\alpha_{zz}$ (see~\eqref{eq:a_nl_ansio}). Therefore, we will consider a general classification of the nonlinear solutions in the core according to the sign of the nonlinear parameter $a^{\texttt{\tiny{EJEM}}}_{nl}$.
\subsubsection {Case \pmb{$a^{\texttt{\tiny{EJEM}}}_{nl}<0$} } 
\label{subsec:negative_anl}
In this case, we set $a^{\texttt{\tiny{EJEM}}}_{nl}=-|a^{\texttt{\tiny {EJEM}}}_{nl}|$ in~\eqref{eq:C0_cont_conditions}  to write the integration constant in the nonlinear core as
\begin{align}
\label{eq:C0_cont_conditions_negative_anl}
C_{0}=k_{0}^{2}H_{[-{d_{core}}/{2}]}\left[\left( \frac{\epsilon_{zz}}{\epsilon_{m}} \right)^{2}q_{m}^{2}-q_{core}^2-\frac{|a^{\texttt{\tiny {EJEM}}}_{nl}|}{2}H^{2}_{[-{d_{core}}/{2}]}\right].
\end{align}
Here, we cannot determine directly the sign of $C_{0}$ according to the sign of $q^2_{core}$, hence the magnetic field $H_{y}$ will be classified according to the signs of $q^2_{core}$ and $C_{0}$. Nevertheless, we have found that the only bounded solutions in the nonlinear core in case of a negative value of $a^{\texttt{\tiny{EJEM}}}_{nl}$ can be obtained only when $q^2_{core}<0$ and $C_{0}>0$ with the following criterion  
\begin{align}
\label{eq:condition_real_Hy_caseIa}
q_{core}^{4} \geq \frac{2 C_{0} |a^{\texttt{\tiny {EJEM}}}_{nl}|}{k_{0}^2}.
\end{align}
In order to clarify this point, we write~\eqref{eq:first_integral_nl_wave_eq_FBM_implicit} in the reduced from
\begin{align}\label{eq:first_integral_negative_a_nl_negative_q_sqr_core_pos_C0} 
\frac{d H_{y}}{\sqrt{A C_{0}-AQ H_{y}^{2}+H_{y}^4(x)}}=\pm \sqrt{\frac{1}{A}}dx,
\end{align}
where the reduced parameters are given by
\begin{align}
\label{eq:A_Q_caseIa}
A={2}/ \left( {k_{0}^2|a^{\texttt{\tiny {EJEM}}}_{nl}|} \right), ~~~~~
Q=k_{0}^2 |q^{2}_{core}|. 
\end{align}
Now,~\eqref{eq:first_integral_negative_a_nl_negative_q_sqr_core_pos_C0} can be written as
\begin{align}\label{eq:first_integral_negative_a_nl_negative_q_sqr_core_pos_C0_delta_gamma} 
\frac{d H_{y}}{\sqrt{\left( \gamma^2-H_{y}^{2}\right)\left( \delta^2-H_{y}^{2}\right)}}=\pm \sqrt{\frac{1}{A}}dx,
\end{align}
with
\begin{equation}
\begin{aligned}
\label{eq:gamma_delta_sqr_caseIa}
\gamma^2=\left( {AQ+\sqrt{(AQ)^2-4AC_{0}}}\right)/2,\\
\delta^2=\left( {AQ-\sqrt{(AQ)^2-4AC_{0}}} \right) /2, 
\end{aligned}
\end{equation}
where $\gamma^2 > \delta^2$. 
 We are looking for real values of the magnetic field $H_{y}$, thus $\gamma^2$ and $\delta^2$ must be real, which means that we need to consider the case $(AQ)^2 \geq 4AC_{0}$, this gives the condition shown in~\eqref{eq:condition_real_Hy_caseIa}. In order to ensure that the quantity under the square root in the denominator of the left-hand side of~\eqref{eq:first_integral_negative_a_nl_negative_q_sqr_core_pos_C0_delta_gamma} is positive, we must have $H^{2}_{y} < \delta^{2}<\gamma^{2}$. It is worth mentioning that we have found an unbounded solution in the core when the condition in~\eqref{eq:condition_real_Hy_caseIa} is not satisfied (see appendix~\ref{appendixI:unbounded_solutions} for more details). 
Integrating ~\eqref{eq:first_integral_negative_a_nl_negative_q_sqr_core_pos_C0_delta_gamma} in the nonlinear core and using formula 17.4.45 from~\cite{Abramowitz-Handbook-Mathematical-functions}, we can express the magnetic field profile $H_{y}$ in the nonlinear core (for $a^{\texttt{\tiny {EJEM}}}_{nl}<0$, $q^2_{core}<0$, and $C_{0}>0$) in terms of the bounded Jacobi elliptical function sn$[u|m]$ with argument $u$ and parameter $m$~\cite{Abramowitz-Handbook-Mathematical-functions,Byrd54-Handbook-elliptic-integrals,Salas14-duff-eq,David-mathematica07} as:
\begin{equation}
\label{eq:Hy_caseIa_negative_anl_v2}
 H_{y}(x)= 
\left\{
\begin{array}{ll}
        \left. \delta \text{sn} \left[ - \sqrt{\frac{\gamma^2}{A} } \left(x+\frac{d_{core}}{2} \right)+X_{0}   \right\vert  \text{m}  \right] ~~~~\text{for}~~~~ \epsilon_{zz}>0 ,	\\ \\
     \left. \delta \text{sn} \left[ + \sqrt{\frac{\gamma^2}{A} } \left(x+\frac{d_{core}}{2} \right)+X_{0}   \right\vert  \text{m}\right]~~~~\text{for}~~~~~~ \epsilon_{zz}<0.      
\end{array} 
\right. 
\end{equation}
Where,
\begin{equation}
\begin{aligned}
\label{eq:gamma_delta_sqr_caseIa}
\text{m}=\frac{\delta^2}{\gamma^2}~~ \text{and}
\left. X_{0}= \text{sn}^{-1}\left[  \frac{H_{[-{d_{core}}/{2}]}}{\delta} \right\vert \text{m} \right].
\end{aligned}
\end{equation}
Here, sn$^{-1}[u|m]$ is the inverse Jacobi elliptical function sn$[u|m]$~\cite{Abramowitz-Handbook-Mathematical-functions}. In order to demonstrate the choice of the sign in ~\eqref{eq:Hy_caseIa_negative_anl_v2}, we need to look at~\eqref{eq:first_integral_negative_a_nl_negative_q_sqr_core_pos_C0}. If the condition given by~\eqref{eq:condition_real_Hy_caseIa} is satisfied, the denominator of~\eqref{eq:first_integral_negative_a_nl_negative_q_sqr_core_pos_C0} is positive and the sign in the right-hand side depends only on the infinitesimal change of the magnetic field $dH_{y}$. In order to be able to study the sign of the magnetic field derivative, we consider the continuity condition of the longitudinal component $E_{z}$ at the left core interface
\begin{align}
\label{eq:dHy_right_interface_caseIa_dispersion}
 \left. \frac{-k_{0} q_{m}}{\epsilon_{m}} H_{y} \right\vert_{x=(-{d_{core}}/{2})^{-}} =  \frac{1}{\epsilon_{zz}} \left[ \frac{d H_{y}}{d x}\right]_{x=(-{d_{core}}/{2})^{+}}.
\end{align}
The real part of the permittivity in the metal cladding is negative $\epsilon_{m}<0$, and we are looking for solutions with positive attenuation coefficient in the cladding $q_{m}>0$; additionally, we assume that $H_{y}|_{(x=-{d_{core}}/{2})}>0$. Now, since the magnetic field changes its sign at the interface ($E_{z}$ being continuous) the derivative of the magnetic field inside the nonlinear core and near the interface depends on the sign of $\epsilon_{zz}$ such that: for $\epsilon_{zz}<0$ the sign of the magnetic field derivative is positive, while if $\epsilon_{zz}>0$ the correct choice of the magnetic field derivative at the interface is the negative sign. Consequently, according to the sign of $\epsilon_{zz}$, we choose the sign in the right-hand side of~\eqref{eq:first_integral_negative_a_nl_negative_q_sqr_core_pos_C0}. It is important to notice that, the solution shown in~\eqref{eq:Hy_caseIa_negative_anl_v2} for $a^{\texttt{\tiny {EJEM}}}_{nl}<0$, cannot be obtained in the isotropic case with positive Kerr-nonlinearity. Nevertheless, due to the full anisotropic treatment of the effective nonlinearity in the present study, we can obtain a nonlinear solution with $a^{\texttt{\tiny {EJEM}}}_{nl}<0$, starting from a positive Kerr-nonlinearity (see section~\ref{sec:numerical_results} for an example of this case), which can also be seen by looking at~\eqref{eq:a_nl_ansio}. This is one of the consequences of the anisotropic nonlinear core considered in the current study.

The procedure of the derivation of the nonlinear dispersion relation is similar to what have already been used for the isotropic case~\cite{Walasik15a,Walasik15b}: we use the magnetic field profile in the nonlinear core~\eqref{eq:Hy_caseIa_negative_anl_v2}, the magnetic field in the linear metal claddings (\eqref{eq:Hy_linear_cladding}), and Maxwell's equations (equations~\ref{eq:maxwell_TM}). Using the continuity conditions for the tangential electromagnetic field components  $H_{y}$ and $E_{z}$ at the right core interface ($x=d_{core}/2$), we can write the nonlinear dispersion relation in terms of the bounded Jacobi elliptical functions sn$[u,m]$, cn$[u,m]$, and dn$[u,m]$ with argument $u$ and parameter $m$ as
\begin{equation}
\begin{aligned}
\label{eq:dispersion_relation_final_caseIa_epszz_neg}
\left. \frac{-k_{0} \epsilon_{zz} q_{m}}{\epsilon_{m}}   \text{sn}  \left[\pm \sqrt{\frac{\gamma^2}{A}} d_{core}+X_{0} \right\vert \text{m} \right] = ~~~~~~~~~~~~~~~~~~~~~~~~~~~\\
   \left.  \pm \sqrt{\frac{\gamma^2}{A}} \left\lbrace  \text{cn}  \left[\pm \sqrt{\frac{\gamma^2}{A}} d_{core}+X_{0}  \right\vert \text{m} \right] \right\rbrace~~ \\
   \times  \left.  \left\lbrace \text{dn} \left[ \pm \sqrt{\frac{\gamma^2}{A}} d_{core}+X_{0}  \right\vert \text{m} \right]  \right\rbrace.
\end{aligned}
\end{equation}
 In~\eqref{eq:dispersion_relation_final_caseIa_epszz_neg}, the sign in front of the square roots is related to the sign of $\epsilon_{zz}$ as we discussed before (see~\eqref{eq:Hy_caseIa_negative_anl_v2} and the text after). We remind the reader that $\epsilon_{xx}$ appears in \eqref{eq:dispersion_relation_final_caseIa_epszz_neg} through $q^{2}_{core}$ that is used to define $Q$ and consequently $\gamma^{2}$ and $\delta^{2}$. Since we are searching for nonlinear bounded solutions with finite energy in the core, we will consider only the subcase with $q^2_{core}<0$ and $C_{0}>0$ for negative $a^{\texttt{\tiny{EJEM}}}_{nl}$ (which satisfies the condition shown in~\eqref{eq:condition_real_Hy_caseIa}) in the derivation of the nonlinear dispersion relation and the other subcases which provide unbounded solutions in the nonlinear core will be summarized in appendix~\ref{appendixI:unbounded_solutions} and table~\ref{tab:Hy_unbounded_negative_anl}. 
\subsubsection{Case \pmb{$a^{\texttt{\tiny {EJEM}}}_{nl}>0$}} 
\label{subsec:positive_anl}
Now we consider the case in which the effective parameters $\epsilon_{xx}$, $\epsilon_{zz}$, $\alpha_{xx}$, and $\alpha_{zz}$ provide positive values for the effective nonlinearity $a^{\texttt{\tiny {EJEM}}}_{nl}$ (see~\eqref{eq:a_nl_ansio} ). The situation is quite similar to the isotropic case with focusing Kerr-type nonlinearity \cite{Walasik15a,Walasik15b}, while the present anisotropic case is more general since it is not necessary to use focusing Kerr-type nonlinearity to get positive values of $a^{\texttt{\tiny {EJEM}}}_{nl}$, as it can be inferred from~\eqref{eq:a_nl_ansio}. The isotropic case studied previously~\cite{Walasik15a}, can be seen as a special case of the current study. For $a^{\texttt{\tiny {EJEM}}}_{nl}>0$, the nonlinear wave equation given by~\eqref{eq:first_integral_nl_wave_eq_FBM_implicit} can be written as
\begin{equation}\label{eq:first_integral_positive_a_nl} 
\left\{
\begin{aligned}{}
\frac{d H_{y}}{\sqrt{A C_{0}+AQ H_{y}^{2}-H_{y}^4(x)}}=\pm \sqrt{\frac{1}{A}}dx~~~~~\text{for}~~~~~ q^{2}_{core}>0,	\\
\frac{d H_{y}}{\sqrt{A C_{0}-AQ H_{y}^{2}-H_{y}^4(x)}}=\pm \sqrt{\frac{1}{A}}dx~~~~~\text{for}~~~~~ q^{2}_{core}<0,      
\end{aligned} 
\right. 
\end{equation}
with the parameters $A$ and $Q$ such that
\begin{equation}
\label{eq:A_Q_caseII_pos_anl}
\begin{aligned}
A=\frac{2}{\left(k^2_{0} a^{\texttt{\tiny {EJEM}}}_{nl}\right)},~  \text{and} ~~~
Q=
\left\{
\begin{array}{ll}
  k_{0}^2 q^{2}_{core}~~~~~~~\text{for}~~~~ q^{2}_{core}>0,	\\
  k_{0}^2 |q^{2}_{core}|~~~~~\text{for}~~~~ q^{2}_{core}<0.
\end{array} 
\right. 
\end{aligned}
\end{equation}
This means that~\eqref{eq:first_integral_positive_a_nl} takes different forms according to the signs of the integration constant $C_{0}$ and the quantity $q^2_{core}$. 

For $q^2_{core}<0$: in this subcase, the integration constant can only take positive value as it can be inferred from~\eqref{eq:C0_cont_conditions} for $H_{[-d_{core}/2]}>0$. In this case, the magnetic field profile $H_{y}(x)$ will be written in terms of the bounded Jacobi elliptical function cn$[u|m]$ as
\begin{equation}
\label{eq:Hy_caseIIa_positive_anl_final}
 H_{y}(x)= 
\left\{
\begin{array}{ll}
        \left. \delta \text{cn} \left[ + \sqrt{\frac{\gamma^2+\delta^2}{A} } \left(x+\frac{d_{core}}{2} \right)+X_{0}   \right\vert  \text{m} \right]  ~~~~\text{for}~ \epsilon_{zz}>0 ,	\\ \\
    \left. \delta \text{cn} \left[ - \sqrt{\frac{\gamma^2+\delta^2}{A} } \left(x+\frac{d_{core}}{2} \right)+X_{0}   \right\vert  \text{m} \right] ~~~~\text{for}~ \epsilon_{zz}<0,      
\end{array} 
\right. 
\end{equation}
where,
\begin{equation}
\begin{aligned}
\label{eq:gamma_delta_sqr_caseIIa}
\gamma^2=\frac{+AQ+\sqrt{(AQ)^2+4AC_{0}}}{2},\\
\delta^2=\frac{-AQ+\sqrt{(AQ)^2+4AC_{0}}}{2}, \\
\text{m}=\frac{\delta^2}{\gamma^2+\delta^2}, \text{and}~~
\left. X_{0}= \text{cn}^{-1}\left[  \frac{H_{[-{d_{core}}/{2}]}}{\delta} \right\vert \text{m} \right].
\end{aligned}
\end{equation}
The choice of the sign in front of the square roots in~\eqref{eq:Hy_caseIIa_positive_anl_final} is related to the sign of $\epsilon_{zz}$ and it is treated as in the previous case $a^{\texttt{\tiny {EJEM}}}_{nl}<0$ (see~\eqref{eq:dHy_right_interface_caseIa_dispersion} and the text after together with~\eqref{eq:first_integral_positive_a_nl}). In order to derive the nonlinear dispersion relation, in this case, we use the magnetic field profile in the nonlinear core~(\eqref{eq:Hy_caseIIa_positive_anl_final}) and the field profile in the linear metal cladding~(\eqref{eq:Hy_linear_cladding}) together with the continuity condition of the tangential components of the electromagnetic field at the right core interface. The final expression of the nonlinear dispersion relation reads
\begin{equation}
\begin{aligned}
\label{eq:dispersion_caseIIa_dispersion_pos_neg_epszz_qsqr_neg_C0_pos}
 \left. \frac{-k_{0} \epsilon_{zz} q_{m}}{\epsilon_{m}} \text{cn} \left[ \mp \sqrt{\frac{\gamma^2+\delta^2}{A} }   d_{core} +X_{0}   \right\vert  \text{m} \right]~~~~~~~~~~~~
 \\ \left.  \mp  \sqrt{\frac{\gamma^2+\delta^2}{A} } \left\lbrace  \text{sn} \left[ \mp \sqrt{\frac{\gamma^2+\delta^2}{A} } d_{core} +X_{0}   \right\vert  \text{m} \right] \right\rbrace\\
 \times \left. \left\lbrace \text{dn} \left[ \mp \sqrt{\frac{\gamma^2+\delta^2}{A} } d_{core} +X_{0}   \right\vert  \text{m} \right] \right\rbrace=0.
\end{aligned}
\end{equation}
We choose the lower positive sign for $\epsilon_{zz}>0$ and the upper negative sign for $\epsilon_{zz}<0$.
\\ \\For $q^2_{core}>0$: unlike the former subcase, the sign of the integration constant $C_{0}$ cannot be determined directly. \\ First, we consider $C_{0}<0$ . Again,  the magnetic field profile in the nonlinear core can be obtained by integrating~\eqref{eq:first_integral_positive_a_nl} using formula 17.4.43 from~\cite{Abramowitz-Handbook-Mathematical-functions} with the reduced parameters $A$ and $Q$ defined in~\eqref{eq:A_Q_caseII_pos_anl} such that
\begin{equation}
\label{eq:Hy_caseIIa_positive_anl_IIb_neg_C0_final}
 H_{y}(x)= 
\left\{
\begin{array}{ll}
        \left. \delta \text{nd} \left[ - \sqrt{\frac{\gamma^2}{A} } \left(x+\frac{d_{core}}{2} \right)+X_{0}   \right\vert  \text{m} \right],   ~~~~\text{for}~~\epsilon_{zz}>0 ,	\\ \\
   \left. \delta \text{nd} \left[ + \sqrt{\frac{\gamma^2}{A} } \left(x+\frac{d_{core}}{2} \right)+X_{0}   \right\vert  \text{m} \right],  ~~~~\text{for}~~ \epsilon_{zz}<0,      
\end{array} 
\right. 
\end{equation}  
where
\begin{equation}
\begin{aligned}
\label{eq:gamma_delta_sqr_caseIIb_neg_c0}
\gamma^2=\frac{AQ+\sqrt{(AQ)^2-4A|C_{0}|}}{2},\\
\delta^2=\frac{AQ-\sqrt{(AQ)^2-4A|C_{0}|}}{2}, \\
\text{m}=\frac{\gamma^2-\delta^2}{\gamma^2}, \text{and}~~
\left. X_{0}= \text{nd}^{-1}\left[  \frac{H_{[-{d_{core}}/{2}]}}{\delta} \right\vert \text{m} \right].
\end{aligned}
\end{equation}
It is worth mentioning that according to the parameters used, the quantities under the square roots in the definitions of $\gamma^2$ and $\delta^2$ shown in~\eqref{eq:gamma_delta_sqr_caseIIb_neg_c0} are positive which ensures that both $\gamma$ and $\delta$ are real quantities and the magnetic field profile provided by~\eqref{eq:Hy_caseIIa_positive_anl_IIb_neg_C0_final} is real. Using the magnetic field profile shown in~\eqref{eq:Hy_caseIIa_positive_anl_IIb_neg_C0_final} and proceeding as for the previous case $a^{\texttt{\tiny {EJEM}}}_{nl}<0$, we can write the nonlinear dispersion relation as
\begin{equation}
\begin{aligned}
\label{eq:dispersion_caseIIb_negative_C0_neg_pos_epszz}
 \left. \frac{-k_{0} \epsilon_{zz} q_{m}}{\epsilon_{m}} \text{nd} \left[ \mp \sqrt{\frac{\gamma^2}{A} } {d_{core}} +X_{0}   \right\vert  \text{m} \right]= ~~~~~~~~~~~~~~~~~~~~~~~~~~~~~~~~\\
 \left. \mp  \sqrt{\frac{\gamma^2}{A} }~(\text{m}) \left\lbrace \text{sd} \left[ \mp \sqrt{\frac{\gamma^2}{A} } {d_{core}} +X_{0}   \right\vert  \text{m} \right] \right\rbrace ~~\\
\times  \left.   \left\lbrace \text{cd} \left[ \mp \sqrt{\frac{\gamma^2}{A} } {d_{core}} +X_{0}   \right\vert  \text{m} \right] \right\rbrace.
\end{aligned}
\end{equation}
In~\eqref{eq:dispersion_caseIIb_negative_C0_neg_pos_epszz} we choose the upper negative sign in front of the square roots for $\epsilon_{zz}>0$, and we choose the bottom positive sign for $\epsilon_{zz}<0$~(see \eqref{eq:dHy_right_interface_caseIa_dispersion}, and the derivative of the Jacobi elliptical functions nd$[u,m]$ in~\cite{Abramowitz-Handbook-Mathematical-functions}).
\\ \\Second, for $C_{0}>0$, we can write the magnetic field profile in the nonlinear core as:
\begin{equation}
\begin{aligned}
\label{eq:Hy_caseIIb_positive_pos_C0_final}
 H_{y}(x)=~~~~~~~~~~~~~~~~~~~~~~~~~~~~~~~~~~~~~~~~~~~~~~~~~~~~~~~~~~~~~~~~~~~~~~~~~~~~~~~~~~~~~
\\ \left\{
\begin{array}{ll}
        \left. \frac{\gamma \delta}{\sqrt{\gamma^2+\delta^2}} \text{sd} \left[ - \sqrt{\frac{\gamma^2+\delta^2}{A} } \left(x+\frac{d_{core}}{2} \right)+X_{0}   \right\vert  \text{m} \right]   ~~\text{for}~ \epsilon_{zz}>0,	\\ \\
   \left. \frac{\gamma \delta}{\sqrt{\gamma^2+\delta^2}} \text{sd} \left[ + \sqrt{\frac{\gamma^2+\delta^2}{A} } \left(x+\frac{d_{core}}{2} \right)+X_{0}   \right\vert  \text{m} \right]  ~~\text{for}~ \epsilon_{zz}<0,      
\end{array} 
\right. 
\end{aligned}
\end{equation} 
where
\begin{equation}
\begin{aligned}
\label{eq:gamma_delta_sqr_caseIIb_pos_c0}
\gamma^2=\frac{-AQ+\sqrt{(AQ)^2+4AC_{0}}}{2},~~~~~~~~~~~~~~~~~~~~~~~~~~~~~~~\\
\delta^2=\frac{+AQ+\sqrt{(AQ)^2+4AC_{0}}}{2},~~~~~~~~~~~~~~~~~~~~~~~~~~~~~~~ \\
\text{m}=\frac{\delta^2}{\gamma^2+\delta^2}, \text{and}~~
\left. X_{0}= \text{sd}^{-1}\left[  \frac{H_{[x=-{d_{core}}/{2}]}\sqrt{\gamma^2+\delta^2} }{\gamma \delta} \right\vert \text{m} \right].
\end{aligned}
\end{equation}
The associated nonlinear dispersion relation gives
\begin{equation}
	\begin{aligned}
		\label{eq:dispersion_caseIIb_positive_C0_final_neg_pos_epszz} 
		\left. \frac{-k_{0} \epsilon_{zz} q_{m}}{\epsilon_{m}\sqrt{\gamma^2+\delta^2}} \left\lbrace \text{sn} \left[ \mp \sqrt{\frac{\gamma^2+\delta^2}{A} } d_{core}+X_{0}   \right\vert  \text{m} \right]  \right\rbrace \\ \times
		\left.  \left\lbrace  \text{dn} \left[ \mp \sqrt{\frac{\gamma^2+\delta^2}{A} } d_{core}+X_{0}   \right\vert  \text{m} \right]   \right\rbrace \\
		\left. =   \mp \sqrt{\frac{1}{A} }   \text{cn} \left[ \mp \sqrt{\frac{\gamma^2+\delta^2}{A} } d_{core}+X_{0}  \right\vert  \text{m} \right]. 
	\end{aligned}
\end{equation}
It is important noticing that an alternative formula  to~\eqref{eq:Hy_caseIIb_positive_pos_C0_final} can also be obtained in terms of the Jacobi elliptical function cn$[u|m]$ as it was already used in the isotropic case~\cite{Walasik15a}. 
The choice of the sign of the nonlinear dispersion relation in~\eqref{eq:dispersion_caseIIb_positive_C0_final_neg_pos_epszz} is linked to the sign of $\epsilon_{zz}$ as it is shown in~\eqref{eq:Hy_caseIIb_positive_pos_C0_final} in which we choose the upper negative sign for $\epsilon_{zz} >0$ and the lower positive one for $\epsilon_{zz}<0$.

 The nonlinear dispersion relations depicted in equations~(\ref{eq:dispersion_relation_final_caseIa_epszz_neg}),~(\ref{eq:dispersion_caseIIa_dispersion_pos_neg_epszz_qsqr_neg_C0_pos}),~(\ref{eq:dispersion_caseIIb_negative_C0_neg_pos_epszz}), and~(\ref{eq:dispersion_caseIIb_positive_C0_final_neg_pos_epszz}) represent the full nonlinear dispersion relations for  bounded solutions propagating in  the anisotropic nonlinear plasmonic slot waveguide depicted in figure~\ref{fig:structure} under the EJEM assumptions described in subsection~\ref{subsec:EJEM}. For a fixed opt-geometrical parameters, the above nonlinear dispersion relations are solved for the real part of the effective index $\Re e(n_{eff})$ to obtain the nonlinear dispersion diagrams.
\subsection{Finite-element method}
\label{subsec:FEM}
In this part, we use the finite-element based model (FEM) to solve the  nonlinear TM eigenvalue problem in the one-dimensional anisotropic layered structure depicted in figure~\ref{fig:structure}. It is well-known that the FEM is generally versatile and can be applied to complex nonlinear waveguide problems, including the two-dimensional ones with arbitrary shape and field profiles~\cite{Rahman90,li1992variational,polstyanko1996nonlinear,desevedavy2009Te-As-Se-optical-fiber}. Generally speaking, in the frame of the FEM, the initial physical problem is transformed into a variational form (weak formulation) by multiplying the initial partial differential equation by chosen test functions that belong to a particular functional space. The next step is the discretization of the problem in which the waveguide cross section is first divided into a patchwork of elements. The unknown fields are expanded in terms of interpolation polynomials over each element. The expansion coefficients that define the values of the fields at the nodal elements can then be obtained by solving a standard matrix eigenvalue problem. For a general and recent review of the finite-element method in the frame of optical waveguides, the reader can refer to chapter $4$ in reference~\cite{livre12-FPCF}. 

In this article, in order to treat the anisotropic nonlinearity in the frame of our FEM, the fixed power algorithm~\cite{Rahman90,Ferrando03_fixed_power,Drouart08} will be used, in which the input is the total power and the outputs are the field profile (the eigenfunction) and the corresponding effective index (the eigenvalue). This algorithm uses a simple iterative scheme based on a sequence of linear modal solutions which converge to the nonlinear solution after few steps. The fundamental issue is that the amplitudes of  the eigenmode are irrelevant and the numerical solution of the intermediary eigenvalue problem used at each iteration has an uncontrolled amplitude. But, since the nonlinear problem depends on the amplitude of the field, the numerical eigenvector which is computed at each step as  a solution of the eigenvalue problem has to be scaled by a scalar factor. This process allows the eigenvector to be used as input for the next iteration. The scaling factor is computed at each iteration for a fixed value of the input power. 
For the single-component eigenvalue problem in terms of the magnetic field component $H_{y}$, an approximated formula  has already been used~\cite{Drouart08,Walasik14,Elsawy_OL_16,Elsawyspie15}  in order to compute the scaling factor using only the linear part of the permittivity and neglecting the nonlinear term (see~\eqref{eq:chi_Ex_lin_epsxx} in appendix~\ref{appendixII:adapted_FEM}). Moreover, only the transverse component of the electric field was used in the isotropic Kerr-type nonlinearity. It is worth mentioning, that it is not possible to take into account all the electric field components in the Kerr-type nonlinearity using the single-component eigenvalue problem~\cite{Walasik14} within the FEM implementation of fixed power algorithm, as it is demonstrated in~\eqref{eqn:chi_Ex_full_single_wave_eq} and the text after it.

In the current study, we present a new and generalized formalism for the fixed power algorithm in arbitrary nonlinear layered structures where the nonlinearity will be treated in a more rigorous way such that all the components of the electric field are taken into account in the Kerr-type nonlinearity and no assumptions are needed to compute the scaling factor. Additionally, unlike the previous isotropic cases~\cite{Rahman90,Drouart08,Walasik14,Elsawy_OL_16}, we consider a fully anisotropic nonlinear treatment for the permittivity tensor as it is shown in \eqref{eq:real_permitt_tensor}. 

Our approach is based on the solution of a coupled nonlinear eigenvalue problem in terms of the continuous tangential components of the electromagnetic field $H_{y}$ and $E_{z}$. In order to obtain the correct weak formulation, we must consider the full TM wave equations with both the inhomogeneous permittivity term induced by the nonlinearity and the structure interfaces. The coupled weak formulation reads:
\begin{equation}\label{eq:weak_formulation_Hy_Ez_v3}
\left\{\begin{aligned}
          \frac{-1}{k^2_{0}}\int_{\Gamma}  \frac{1}{\epsilon_{z}(x)} \frac{d h_{y}}{d x} \frac{d h^{\prime}_{y}}{d x} dx +\int_{\Gamma} h_{y}(x) h^\prime_{y}(x) dx ~~~~~~~~~~~~~~~~~~~~~~~~~~~~~~~~~~~~~~~~~~~~~~~~~~ \\ 
          -  n_{eff}^2 \int_{\Gamma} \frac{1}{\epsilon_{x}(x)} h_{y}(x) h^\prime_{y}(x)dx=0~~~~~~~~~~~~~~~~~~~~~~, \\ 
          \int_{\Gamma} e_{z}(x)e^{\prime}_{z}(x) dx- \frac{1}{\epsilon_{0}k_{0}c}\int_{\Gamma}  \frac{1}{\epsilon_{z}(x)} \frac{d h_{y}}{d x} {e^{\prime}_{z}}  dx =0.~~~~~~~~~~~~~~~~~~~~~~~~~~~~~~~~~~~~~~~~~~~~~
\end{aligned}
\right.
\end{equation}
	Here, $h_{y}(x)$ and $e_{z}(x)$ stand for the tangential component of the magnetic and electric field, respectively, and $\Gamma$ is the domain of integration (in the present case the full cross section of the waveguide). In~\eqref{eq:weak_formulation_Hy_Ez_v3}, $\forall$ $h^\prime_{y}(x), e^\prime_{z}(x) \in \mathtt{H}_{0}^{2}(\Gamma)$ we look for $h_{y}(x), e_{z}(x)   \in \mathtt{H}_{0}^{2}(\Gamma)$, where $\mathtt{H}_{0}^{2}(\Gamma)$ is Sobolev space of order $2$ with null Dirichlet boundary conditions on the domain of integration $\Gamma$. It is important to point out that for the magnetic field  $h_{y}$, we need to pick a test function from a functional space (Sobolev space) at least of order $2$, while for the tangential component of the electric field $e_{z}$ we can choose a test function from a functional space of order $1$ or $2$ as it can be understood from the relation between the tangential components shown in~\eqref{eq:Ez_full}. We use the fixed power algorithm described in algorithm~\ref{alg: fixed_power_full_nl} to solve the coupled eigenvalue problem shown in~\eqref{eq:weak_formulation_Hy_Ez_v3}. Our FEM is implemented using the free and open-source software \texttt{GMSH} as a mesh generator and \texttt{GETDP} as a solver~\cite{dular-getdp-1998general,geuzaine2007getdp,geuzaine2009gmsh}. These software programs have already been used to solve both one-dimensional and two-dimensional isotropic nonlinear electromagnetic waveguide problems~\cite{Drouart08,Elsawy_OL_16,Walasik14}.  It is worth mentioning that our FEM with its fixed power algorithm shown in the algorithm~\ref{alg: fixed_power_full_nl} can be reduced to take into account all EJEM assumptions as described in appendix~\ref{appendixII:adapted_FEM}. 
\begin{algorithm}[H]
\caption{Fixed power algorithm to solve the coupled nonlinear eigenvalue problem depicted in~\eqref{eq:weak_formulation_Hy_Ez_v3}}\label{alg: fixed_power_full_nl}
\begin{algorithmic}[1]
\State We start with an initial guess $E^{init}_{x}$, $E^{init}_{z}$, which will be used to compute $\epsilon_{x}$ and $\epsilon_{z}$ from~\eqref{eq:real_permitt_tensor}.
\State  We use $\epsilon_{x}$ and $\epsilon_{z}$ in the coupled eigenvalue problem~\eqref{eq:weak_formulation_Hy_Ez_v3} to compute the eigenvectors $h_{y}$ and $e_{z}$ with the corresponding eigenvalue $ n_{eff}$. The outputs will be used to compute the rescaling factor $\chi$, for a given fixed value of the power $P_{tot}$ such that  $$ P_{tot} ~= \frac{\chi^{2}}{2k_{0}\Re e(n_{eff})} \int_{\Gamma} \left[ \frac{d e_{z}(x)}{d x}+ \omega \mu_{0}h_{y}(x) \right] h_{y}(x) dx.$$
\State The rescaling factor $\chi$ will be used to compute the correct amplitude of the longitudinal components such that $H_{y}=\chi h_{y}$ and $E_{z}=\chi e_{z}$. The transverse component of the electric field $E_{x}$  will be computed from~\eqref{eq:Ex_Ez_Hy} using $H_{y}$ and $E_{z}$. The updated effective nonlinear permittivities $\epsilon_{x}$ and $\epsilon_{z}$ will be used as inputs for the next iteration.
\State We repeat steps (2) and (3) until the following criterion is satisfied: ${|\Re e(n^{i}_{eff})-\Re e(n^{i-1}_{eff})|}/{|\Re e(n^{i}_{eff})|} < \delta~~\forall i \in [1,N],$
where $\Re e(n^{i}_{eff})$ is the eigenvalue for the step $i$ and N is the step number in the procedure. We set $\delta=10^{-5}$ such that in order to fulfill the criterion between 10 and 15 steps are needed depending on the waveguide parameters and the initial field used.
\end{algorithmic}
\end{algorithm}
\section{Numerical results}
\label{sec:numerical_results}
In this section, we present several numerical results to validate our models and to demonstrate the influence of the anisotropy on the nonlinear dispersion diagrams and on the effective nonlinearity. We build nonlinear cores from stacks of realistic bulk materials, and we use the effective medium theory to retrieve their linear and the nonlinear effective parameters~\cite{ciattoni2010extreme,ciattoni2011all}. Moreover, we assume that both $d_{1}$ and $d_{2}$ (see figure~\ref{fig:structure}(b)) are much smaller than the operating wavelength, so we can derive the effective linear permittivities and the effective nonlinear susceptibilities in the frame of the simple effective medium theory (EMT)~\cite{ciattoni2010extreme,ciattoni2011all}
\begin{equation}
\label{eq:EMT}
\begin{aligned}
\tilde{\epsilon}_{zz}= r\epsilon_{2}+(1-r)\epsilon_{1}, ~~~~~~~~~~~~~~~~~~\tilde{\epsilon}_{xx}=  \frac{{\epsilon_{1}\epsilon_{2}} } {  {r\epsilon_{1}+(1-r)\epsilon_{2}} },~~~~~~~~~~~~~~~~\\
\tilde{\alpha}_{zz}=3(r\chi^{(3)}_{2}+(1-r)\chi^{(3)}_{1}), ~~~~~~
\tilde{\alpha}_{xx}=3\frac {r\chi^{(3)}_{2}\epsilon_{1}^{2}+(1-r)\chi^{(3)}_{1}\epsilon_{2}^{2} } { ((1-r)\epsilon_{2}+r\epsilon_{1})^{2} }, 
\end{aligned}
\end{equation}
where, $r={d_{2}}/  \left( {d_{1}+d_{2}} \right)$ is the ratio of the second material in the core, $\epsilon_{j}$ and $\chi^{(3)}_{j}$, $j \in \{ 1,2  \}$, are the linear permittivities and the third order nonlinear susceptibilities of the constituent materials in the core, respectively. Generally, $\tilde{\epsilon}_{jj}=\epsilon_{jj}+i\epsilon^{\prime \prime}_{jj}$ and  $\tilde{\alpha}_{jj}=\alpha_{jj}+i\alpha^{\prime \prime}_{jj}$ are complex, the real parts are used in the derivation of the models whereas the imaginary parts will be used to estimate the losses. In this study we will consider only the linear losses, nonlinear processes like two-photon absorption are neglected~\cite{Olivier14_renversez_nl_properties_chalco,boyd_nonlinear_2008}. The imaginary part of the effective indices $\Im m(n_{eff})$ will be estimated using the method based on the field profiles and imaginary parts of the permittivity described in~\cite{Davoyan09,Walasik14,Stegeman85}. In formula~(\ref{eq:loss_snyder_aniso}), we provide the extension of this method to the anisotropic case:
 \begin{equation}
 \label{eq:loss_snyder_aniso}
 \Im m(n_{eff})=\frac{\epsilon_0c}{4P_{tot}} \left[\int_{\text{core}}{\epsilon^{\prime \prime}_{xx} |E_{x}(x)|^{2}+\epsilon^{\prime \prime}_{zz} |E_{z}(x)|^{2} dx} + \int_{\text{cladding}}{\epsilon^{\prime \prime}_{m} \left( |E_{x}(x)|^{2}+ |E_{z}(x)|^{2} \right) dx}   \right]
 \end{equation}		
		where, $P_{tot}$ is the total power and $\epsilon^{\prime \prime}_{m}$, being the imaginary part of the metal cladding permittivity. It is worth noting that, this method has been already used to estimate the losses in isotropic nonlinear plasmonic waveguides~\cite{Davoyan09,Walasik14,Elsawy_OL_16}. For the numerical results presented below, the core thickness is fixed at $d_{core}=400$ nm, and we use gold for the metal claddings with $\tilde{\epsilon}_{m}=-90+i10$ at $\lambda=1.55$ $\mu$m~\cite{palik1998handbook}. First, we simplify our models to the isotropic case and compare the results with previously published works. For the anisotropic case, we use the general classification shown in subsection~\ref{subsec:EJEM} for $a^{\texttt{\tiny {EJEM}}}_{nl}<0$ and $a^{\texttt{\tiny {EJEM}}}_{nl}>0$ depending on the values and the signs of the linear and the nonlinear permittivity terms (see~\eqref{eq:a_nl_ansio}). It is worth noticing that this classification is more general than the one shown in our recent work~\cite{Elsawy16-OL-anisotropic-metamaterial-waveguides} where $\alpha_{xx}=\alpha_{zz} >0$ and $\epsilon_{xx} \neq \epsilon_{zz}$  in which we classified the problem according to the signs of the linear permittivity tensor terms into  elliptical and hyperbolic cases only. This previous splitting can be seen as a special case of the current more classification.
\subsection{Isotropic case: validation and comparison with~\cite{Walasik15a,Walasik15b} }
\label{subsubsec:valid_isotropic}
\begin{figure}[htbp]
\centering
\includegraphics[width=\columnwidth,angle=0,clip=true,trim= 0 0 0 0]{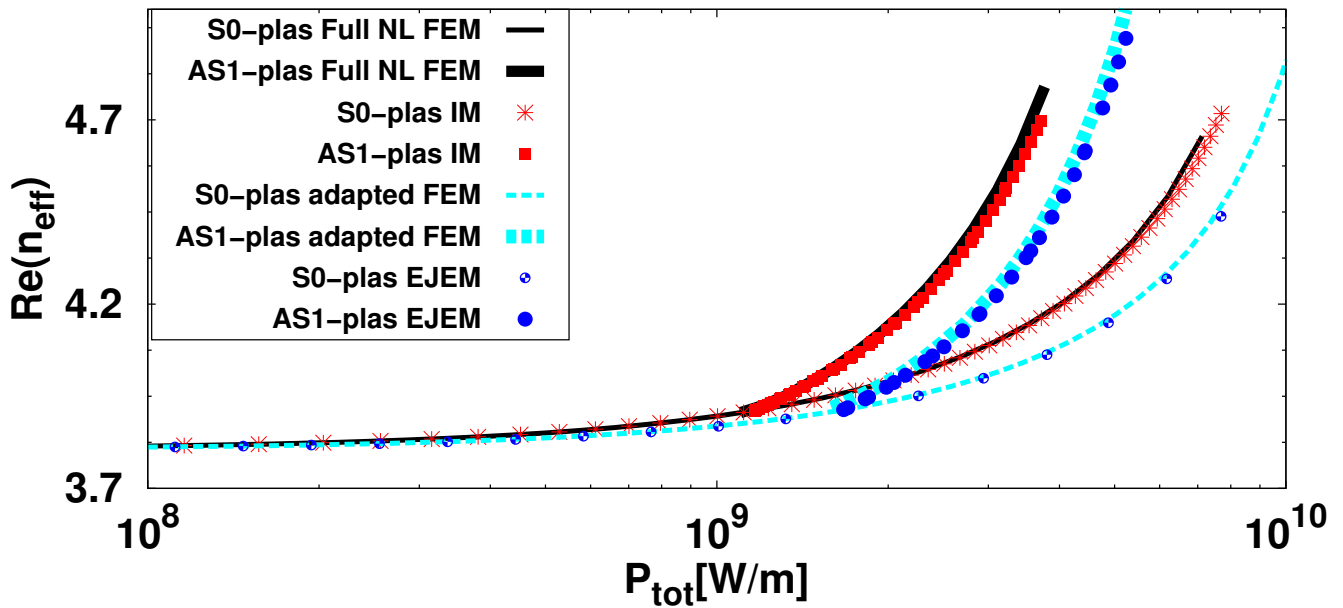}
\caption{Comparison between the nonlinear dispersion curves for the NPSW depicted in figure~\ref{fig:structure} reduced to the fully isotropic case by setting $r=0$ in~\eqref{eq:EMT}  with $\epsilon_{1}=3.46^2+{i}10^{-4}$ and $\alpha_{zz}=\alpha_{xx}=6.36 \times 10^{-19}$ $m^{2}/V^{2}$~\cite{Walasik15b}.}
\label{fig:nonlinear_dispersion_curves_isotropic}
\end{figure}

We begin with the isotropic case, we consider a nonlinear isotropic core with focusing Kerr-type nonlinearity and compare the results with~\cite{Walasik15a,Walasik15b} using exactly the same parameters.  
In this case, we set $r=0$ in~\eqref{eq:EMT}, and the isotropic nonlinear material in the core corresponds to  amorphous hydrogenated silicon with $\epsilon_{1}=3.46^2+{i}10^{-4}$ with nonlinear parameter being $\alpha_{zz}=\alpha_{xx}=6.36 \times 10^{-19}$ $m^{2}/V^{2}$ which is related to the nonlinear refractive index $n_{2}$  through $n_{2} \approx \alpha_{xx}/(\epsilon_{0}c \Re e(\epsilon_{1}))$~\cite{Walasik15a,Walasik15b}. In figure~\ref{fig:nonlinear_dispersion_curves_isotropic}, we compare the nonlinear dispersion curves for the fundamental nonlinear symmetric mode denoted by S0-plas (thin lines and small points) and the first nonlinear asymmetric mode AS1-mode (thick lines and large points), which bifurcates from the S0-plas mode at the critical power value. The asymmetric mode AS1-mode has no counterpart in the linear case and it exists only above a certain threshold as a signature of a strong spatial nonlinear effect~\cite{Akhmediev97,rukhlenko2011dispersion,Walasik15b}. This kind of bifurcation is called Hopf bifurcation and it has already been observed in the simple nonlinear plasmonic slot waveguide~\cite{Davoyan08,Walasik15a,Walasik15b} and in its improved version~\cite{Elsawy_OL_16,Elsawyspie15}. We begin with comparing the results obtained from the FEM described in subsection~\ref{subsec:FEM} and denoted Full NL FEM in the following taking into account all the contributions of the electric field in the Kerr-nonlinearity (black curves) with the results obtained from the interface model (IM) (represented by the red points in figure~\ref{fig:nonlinear_dispersion_curves_isotropic})~\cite{Walasik15a}. The IM is a numerical method developed specifically for nonlinear isotropic slot waveguides~\cite{Walasik15a,Walasik15b} where all the non-null electric field components are present in the Kerr-nonlinearity. One clearly sees that the results obtained from our Full NL FEM with its coupled formulation (see subsection~\ref{subsec:FEM}), agree well with the results acquired using the IM. 
Next, we compare the results obtained from the EJEM-adapted FEM presented in appendix~\ref{appendixII:adapted_FEM} taking into account the assumptions of the EJEM (dashed cyan curves), with the results obtained from the semi-analytical approach EJEM described in subsection~\ref{subsec:EJEM} (blue points). Once again, one clearly sees that the results from our adapted FEM recover the results from the semi-analytical EJEM. In figure~\ref{fig:nonlinear_dispersion_curves_isotropic}, at high powers, there exist some discrepancies between the results  obtained from the methods which consider the full treatment of the nonlinearity (FEM Full NL black curves and IM red curves) in the core and the results obtained in the frame of the EJEM approximations (adapted FEM cyan and EJEM blue points). These discrepancies  are due to the simplified way used to describe the nonlinearity in the EJEM models (see subsection~\ref{subsec:EJEM}). However, the results which are based on the EJEM assumptions can predict the behaviour of the nonlinear dispersion curves as it can be inferred from figure~\ref{fig:nonlinear_dispersion_curves_isotropic}.

Another important remark is that our two FEMs are based on the fixed power algorithm in which, for a given input power, we look for the corresponding effective index of the investigated nonlinear mode (see subsection~\ref{subsec:FEM}), this means that they fail to follow the branches of the nonlinear dispersion curves with a negative slope which usually correspond to unstable modes (see figure 1 in reference~\cite{Walasik15b}). The current versions of the FEM based on the fixed power algorithm, converge only to possibly stable modes~\cite{rahman1991review,li1992variational}. 
\subsection{Anisotropic case with \pmb{$a^{\texttt{\tiny {EJEM}}}_{nl}<0$}: results and validation}
\label{subsec:numerical_example_anl_neg}
\begin{figure}[h]
	\centering
	\includegraphics[width=\columnwidth,angle=0,clip=true,trim= 0 0 0 0]{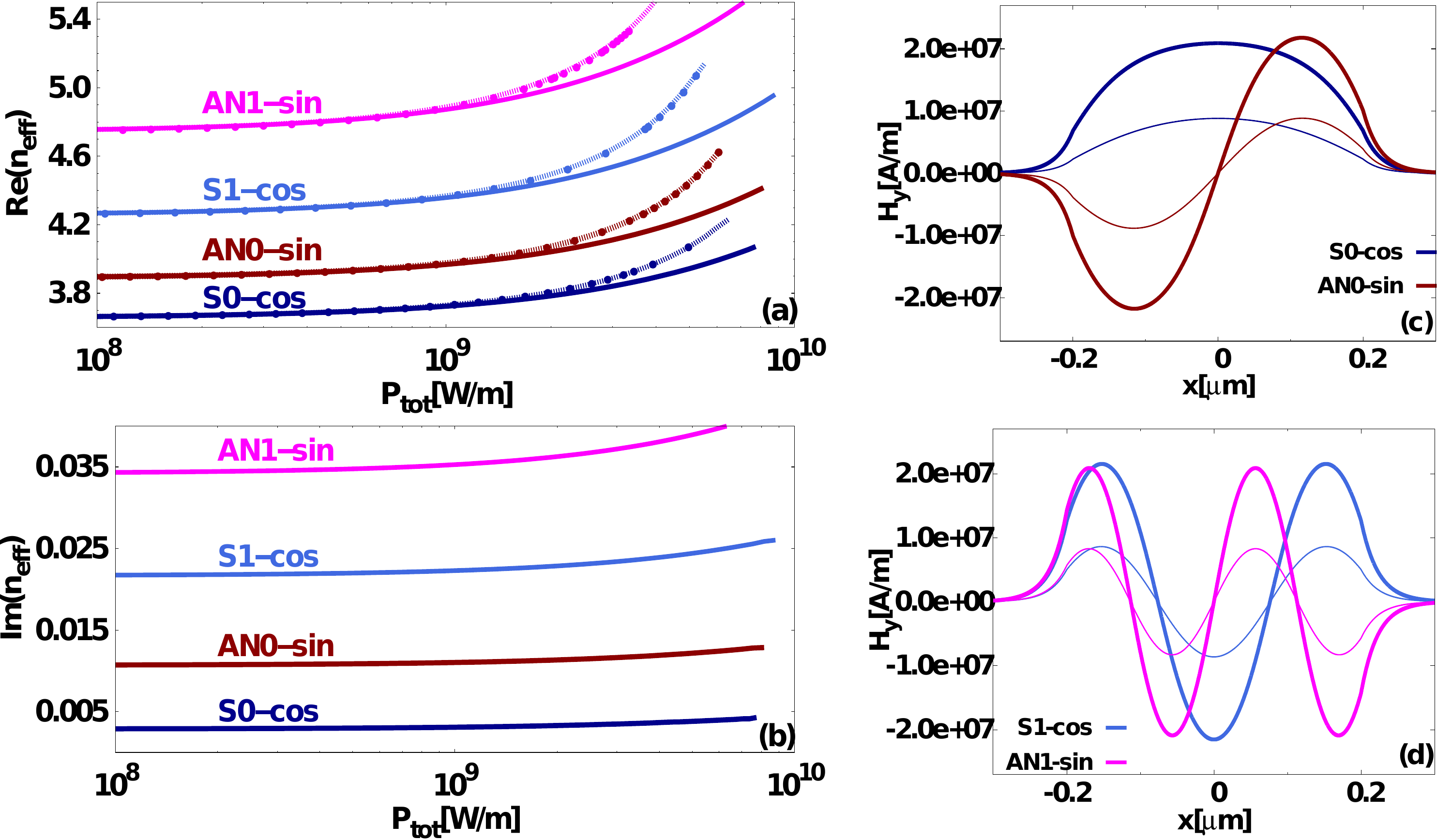}
	\caption{(a) Nonlinear dispersion curves in the anisotropic case for $a^{\texttt{\tiny {EJEM}}}_{nl}<0$ (see the text for the used parameters). Solid lines are for the Full NL FEM, dashed lines for the adapted FEM, and closed circles for the semi-analytical EJEM. (b) $\Im m(n_{eff})$ as a function of $P_{tot}$ for the results obtained with Full NL FEM. (c) and (d) $H_{y}$ component for the modes under investigation at two different power values; thin curves for $P_{tot}=10^{9}$ W/m and thick curves at $P_{tot}=6 \times 10^{9}$ W/m.}
	\label{fig:nl_dispersion_anisotropic_NEGATIVE_anl}
\end{figure} 
In this subsection, we present one example for the case in which the effective nonlinearity is negative (see subsection~\ref{subsec:negative_anl}), and we compare the results obtained from the semi-analytical approach EJEM with those acquired from the Full NL FEM and from the adapted FEM to match the EJEM assumptions (see appendix~\ref{appendixII:adapted_FEM}). As we have mentioned before, the sign of the effective nonlinearity $a^{\texttt{\tiny {EJEM}}}_{nl}$ shown in~\eqref{eq:a_nl_ansio}, depends on the linear and the nonlinear effective parameters of the core, in this example, we choose the two materials in the core in order to get a negative effective nonlinearity. We use silver for material 2 with $\epsilon_{2}=-129+i3.28$~\cite{johnsonPhysRevB1972}. Due to the imaginary part of $\epsilon_{2}$, it is important to use a gain medium for material 1 (with permittivity $\epsilon_{1}$) in order to compensate for the metal losses. The chalcogenide glass is a promising candidate since it can act as a gain medium at the telecommunication wavelength, in addition, it has a high nonlinear coefficient and a low two-photon absorption (TPA) at this wavelength, compared to the usual semiconductors~\cite{nazabal2016luminescence,yan2016Er-Chalco}. Thus, the permittivity of material 1 is set at $\epsilon_{1}=2.47^{2}-i0.0072$ corresponding very small gain. The third order nonlinear susceptibility of material 1 being $\chi^{(3)}_{1}\approx 1.08 \times 10^{-19}$ $m^{2}/V^{2}$~\cite{Olivier14_renversez_nl_properties_chalco}. We will omit the nonlinearity of material 2 (silver) since it is weak compared to the one of the chalcogenide glass at $\lambda=1.55$ $\mu$m. It is worth mentioning that the idea of loss compensation in metal/dielectric multilayer structures has already been investigated theoretically~\cite{argyropoulos13-gain-HMMs,shavelev13-loss-compensation-MMlayered} and demonstrated experimentally~\cite{ciattoni11-ENZ-metamaterial-gain,Grandidier09-PMMA-QDs}. In order to compute the effective parameters in the core, we set $r=0.5$ in~\eqref{eq:EMT}, that gives us, $\tilde{\epsilon}_{xx}=12.80+i10^{-8}$, $\tilde{\epsilon}_{zz}=-61.449+i1.63633$, $\alpha_{xx}=7.15 \times 10^{-19}$ $m^{2}/V^{2}$, and $\alpha_{zz}=1.62 \times 10^{-19}$ $m^{2}/V^{2}$. One clearly sees that the imaginary part of the longitudinal linear component $\epsilon^{\prime \prime}_{zz}$ is not null; in order to fully compensate for the losses in all the directions, we need huge gain values in material 1, which cannot be achieved with the current technology as it is explained in~\cite{smalley14-InGaAsP}. In figure~\ref{fig:nl_dispersion_anisotropic_NEGATIVE_anl}(a), we present the nonlinear dispersion diagram for $Re(n_{eff})$ as a function of the total power using the semi-analytical EJEM with $a^{\texttt{\tiny {EJEM}}}_{nl}<0$ described in subsection~\ref{subsec:negative_anl} (closed circles) and Full NL FEM with the full treatment of the Kerr-nonlinearity (solid lines), and the adapted FEM to match the EJEM assumptions (adapted FEM) shown in appendix~\ref{appendixII:adapted_FEM} (dashed lines).
At low powers, when the assumptions of the EJEM are valid, the dispersion curves acquired by the three methods are identical. Nevertheless, as expected, at high powers, when the EJEM assumptions are no longer valid, we obtain some discrepancies between the Full NL FEM (solid lines) and the methods based on the EJEM assumptions (dashed lines for adapted FEM and closed circles for EJEM). It is worth noticing that the adapted FEM actually reproduces the results obtained from the semi-analytical EJEM, even at high powers, this can be seen as a validation of our methods. Due to the negative effective nonlinearity, we observe a defocusing behaviour inside the nonlinear core, starting from a positive Kerr-nonlinearity; figures~\ref{fig:nl_dispersion_anisotropic_NEGATIVE_anl}(c) (main symmetric and antisymmetric modes) and~\ref{fig:nl_dispersion_anisotropic_NEGATIVE_anl}(d) (first symmetric and antisymmetric higher order modes) illustrate this phenomenon. In figure~\ref{fig:nl_dispersion_anisotropic_NEGATIVE_anl}(b), we estimate the losses 
using the results obtained from the more accurate model Full NL FEM, the  $\Im m(n_{eff})$ is computed using~\eqref{eq:loss_snyder_aniso}, we observe an increase of $\Im m(n_{eff})$ with the increase of the power for the modes under investigation, this is due to the defocusing behaviour  in the core where the field profiles tend to be more located in the metal claddings with the increase of the power (see figures~\ref{fig:nl_dispersion_anisotropic_NEGATIVE_anl}(c) and~\ref{fig:nl_dispersion_anisotropic_NEGATIVE_anl}(d)). It is worth mentioning that in this case, we did not observe bounded asymmetric mode, due to the defocusing  in the core. To conclude on this point, we have presented one example for the case in which the effective nonlinearity is negative, we have validated the methods developed in this study, moreover we have found that, in this example, starting from a positive Kerr-type nonlinearity, we observe a defocusing behaviour  in the core. Different parameters can be chosen in \eqref{eq:a_nl_ansio} in order to obtain a negative effective nonlinearity, nevertheless that do not affect the validation of our models.
 \subsection{Anisotropic case with \pmb{$a^{\texttt{\tiny {EJEM}}}_{nl}>0$}: results and validation}
 \label{subsec:numerical_example_anl_pos}
\begin{figure}[h] 
 \centering
 \includegraphics[width=\columnwidth,angle=0,clip=true,trim= 0 0 0 0 0]{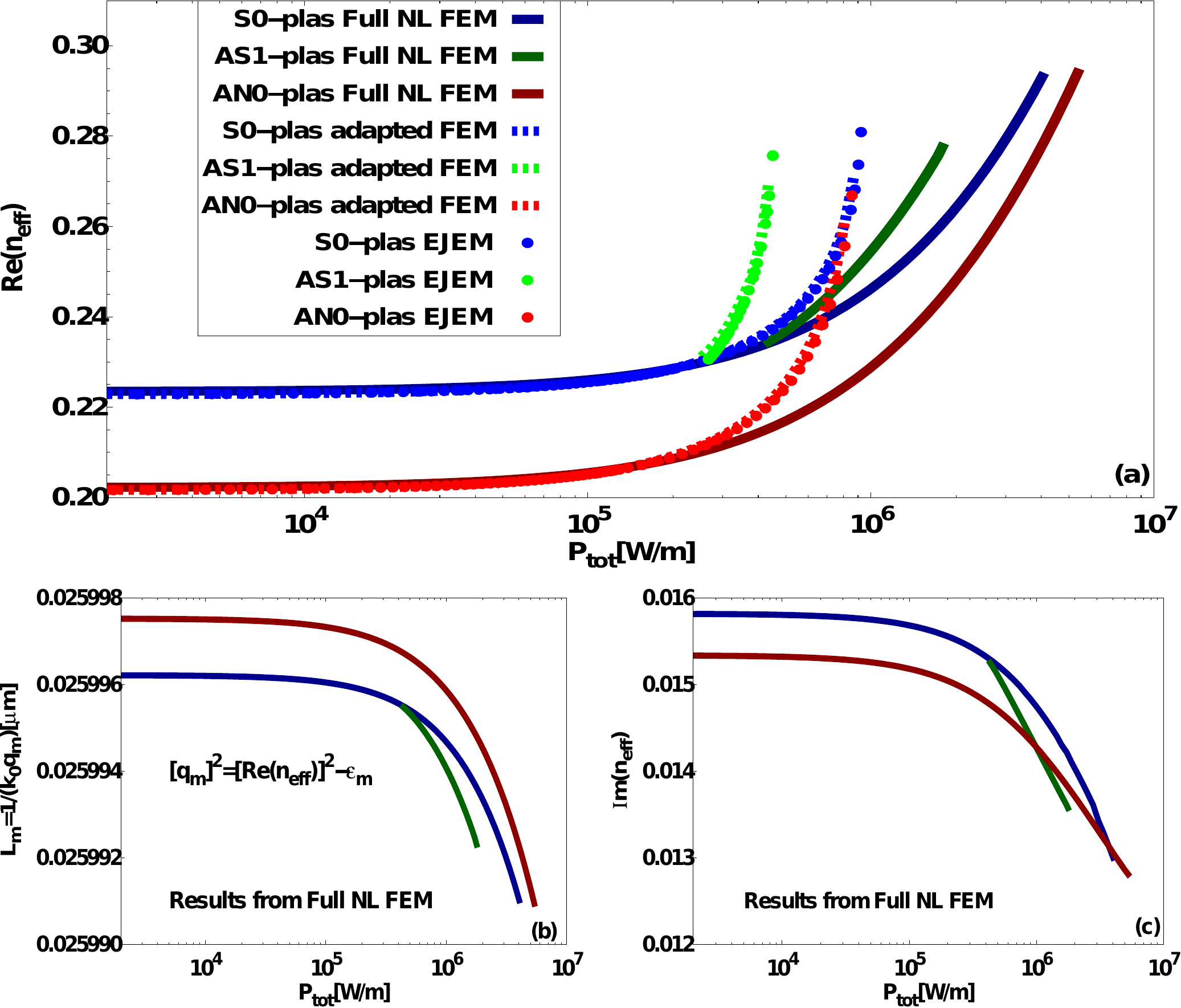}
 \caption{(a): Nonlinear dispersion curves for $a^{\texttt{\tiny {EJEM}}}_{nl}>0$ using EJEM (closed circles), the adapted FEM (dashed lines), and Full NL FEM (dark solid lines). Blue, green, red colours for the symmetric, asymmetric, antisymmetric modes, respectively. (b): The attenuation length in the metal claddings computed using the results obtained from Full NL FEM, and $\epsilon_{m}=-90$ being the real part of the metal cladding permittivity. (c): $\Im m(n_{eff})$ using the results obtained from Full NL FEM, the  corresponding $\Re e(n_{eff})$ are shown in (a). We remind that, in this figure, the effective nonlinear susceptibility parameters $\alpha_{xx}$ and $\alpha_{zz}$ are not equal (see the text for the used parameters).}
 \label{fig:nonlinear_dispersion_curves_anisotropic_anl_pos}
 \end{figure}

\begin{figure}[h] 
 \centering
 \includegraphics[width=\columnwidth,angle=0,clip=true,trim= 0 0 0 0 0]{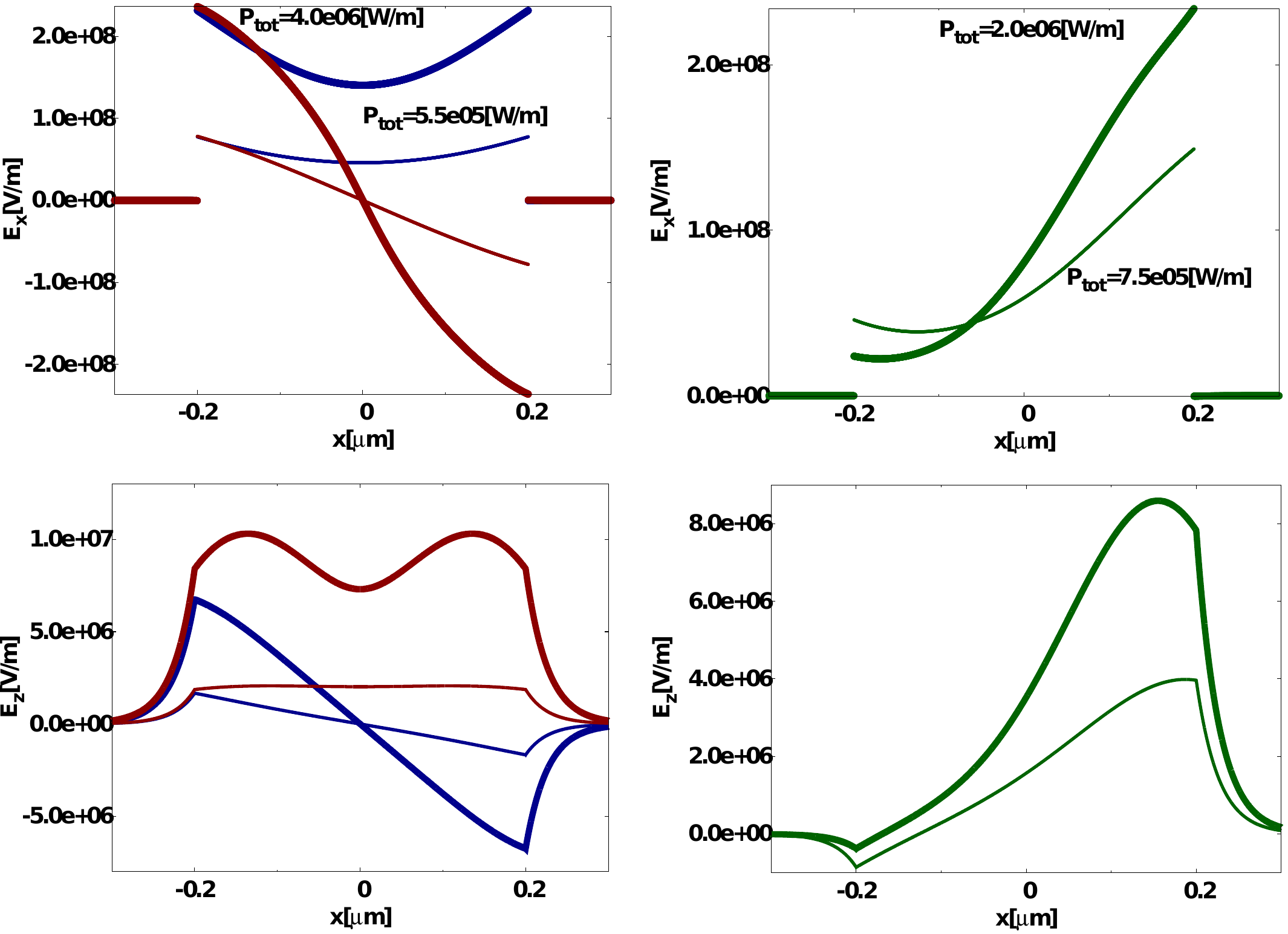}
 \caption{$E_{x}$ (first raw) and $E_{z}$ (second raw) components for the main modes computed using the more accurate model Full NL FEM represented by the solid curves shown in figure~\ref{fig:nonlinear_dispersion_curves_anisotropic_anl_pos}(a). First column: nonlinear symmetric (dark-blue), and antisymmetric (dark-red) modes at two different power values (thin curves at $P_{tot}=5.5e05$ W/m, solid curves at $P_{tot}=4.0e06$ W/m). Second column: nonlinear asymmetric mode at two different power values (thin curves at $P_{tot}=7.5e05$ W/m, solid curves at $P_{tot}=2.0e06$ W/m).}
 \label{fig:Ex_field_profiles_anl_pos}
 \end{figure}

In this subsection, we present a numerical example for the anisotropic case in which $a^{\texttt{\tiny {EJEM}}}_{nl}>0$. We choose material 1 with $\epsilon_{1}=3.46^{2}+i10^{-4}$  and  $\chi^{(3)}_{1}\approx 2.122 \times 10^{-19}$ $m^{2}/V^{2}$ (amorphous hydrogenated silicon~\cite{Walasik15a,Walasik15b}). For the second material in the core with permittivity $\epsilon_{2}$, we consider an ideal ENZ  material with $\Re e(\epsilon_{2}) \approx 0$ and  $\Im m(\epsilon_{2}) \approx 0$. This choice of the second material is justified, since it was recently shown~\cite{Javani2016_ENZ_proerties} that for the usual ENZ materials with $\Re e(\epsilon_{2}) \approx 0$ and $\Im m(\epsilon_{2})>\Re e(\epsilon_{2})$, the fundamental principle of causality leads to diverging energy-loss function and that the loss cannot be compensated even with a huge gain values. A typical example for such usual ENZ materials with null real part and small imaginary part of the permittivity is the indium-tin oxide (ITO)~\cite{Capretti15OL-enhanced-third-harmonic-generation}. Furthermore, it is demonstrated theoretically~\cite{Javani2016_ENZ_proerties} that in order to make use of the properties of the ENZ materials in waveguide problems, they must have a vanishing imaginary part of the permittivity (ideal ENZ). In our study, the ideal ENZ permittivity (material 2 in the core) will be obtained by considering a mixture of aluminium doped zinc oxide (AZO)-coated quantum dots dispersed in a polymer matrix, and the resulting permittivity is retrieved using Maxwell-Garnett's relation. This is technique is similar to what have been used in~\cite{Ciattoni2013IdealENZ}, while here we use spherical AZO-coated quantum dots instead of the spherical Ag-coated quantum dots. The spherical AZO-coated quantum dots in our study consist of AlGaAsP (gain medium with permittivity 11.8-i0.16) as an inner core medium and AZO as an outer shell with permittivity -8+i0.1. Using Maxwell-Garnett's formula~\cite{Ciattoni2013IdealENZ} with a filling ratio set to 0.08, we obtain the permittivity of the spherical AZO-coated quantum dots $\epsilon_{QDs}=-1.42+i0.006$. The next step in obtaining the permittivity of material 2, is to insert the spherical AZO-coated quantum dots within a polymer matrix (PMMA) with $\epsilon_{PMMA}=2.66$ and with a filling ratio equals to 0.475, such that using Maxwell-Garnett's relation, the permittivity of material 2 reads: $\epsilon_{2}=0.0042+i0.000555$. The third order nonlinear coefficient of the second material being $\chi^{(3)}_{2}\approx 3.0 \times 10^{-20}$ $m^{2}/V^{2}$, similar to the one used in~\cite{rizza2011two-peaked-flat-top}. It is worth mentioning that, similar procedure has already been used to obtain an ideal ENZ material using quantum dots~\cite{rizza2011two-peaked-flat-top}. The permittivity of material 2 will be used together with the one of material 1 to obtain the effective parameters in the core using~\eqref{eq:EMT}. We choose $r=0.1$ to obtain  $\tilde{\epsilon}_{xx}=0.042+i0.005$, $\tilde{\epsilon}_{zz}=10.780+i10^{-5}$, $\alpha_{xx}=8.988\times 10^{-19}$$m^{2}/V^{2}$, and  $\alpha_{zz}=5.82\times 10^{-19}$$m^{2}/V^{2}$. In figure~\ref{fig:nonlinear_dispersion_curves_anisotropic_anl_pos}(a), we present the nonlinear dispersion curves for the three main modes: symmetric (blue), asymmetric (green), and antisymmetric (red) obtained using the EJEM (closed circles), and the adapted FEM to match the EJEM assumptions (dashed lines), and with Full NL FEM taking into account the full treatment of the nonlinearity (dark solid curves). 
One can observe that the three methods coincide at low powers, whereas at high power, there exist some discrepancies between them. Once again, this can be understood from the assumptions of the EJEM which are not valid at high powers. In addition, the EJEM (closed circles) and the adapted FEM (dashed lines) are identical even at high powers, which confirms that the discrepancies between the methods at high powers are due to the assumptions of the EJEM. All the methods predict a huge reduction of the bifurcation threshold compared to the isotropic case shown in figure~\ref{fig:nonlinear_dispersion_curves_isotropic}; typically in this anisotropic case, the bifurcation threshold is reduced by more than three orders of magnitude (see green curves in figure~\ref{fig:nonlinear_dispersion_curves_anisotropic_anl_pos}(a)) which is a signature of strong reinforcement of the spatial nonlinearity. This enhancement of the spatial nonlinearity can be understood qualitatively  in~\eqref{eq:a_nl_ansio} for an ENZ $\epsilon_{xx}$ and $\epsilon_{zz}>>1$. For a more detailed study of the influence of the linear anisotropic part of the permittivity tensor with an isotropic nonlinear term unlike the current study, we refer to~\cite{Elsawy16-OL-anisotropic-metamaterial-waveguides}. Another interesting consequences of using an ENZ $\epsilon_{xx}$ and $\epsilon_{zz}>>1$ are the very small effective indices obtained for the modes under investigation (see figure~\ref{fig:nonlinear_dispersion_curves_anisotropic_anl_pos}(a)), which have a direct influence on the field profiles in the metal claddings. This influence can be seen from the attenuation parameter in the metal which is quantified by $q^{2}_{m}=([\Re e(n_{eff})]^{2}-\epsilon_{m})$, where $\epsilon_{m}=-90$ being the real part of the metal cladding permittivity. In this case, $|\epsilon_{m}|>>[\Re e(n_{eff})]^{2}$ even with the increase of $\Re e(n_{eff})$ at high powers, and $q^{2}_{m}$ remains nearly constant. This means that the increase of the power has a negligible effect on the field profiles in the linear metal claddings, and the attenuation length in the metal $L_{m}=1/(k_{0}q_{m})$ remains nearly constant with the increase of the power as it can be seen in figure~\ref{fig:nonlinear_dispersion_curves_anisotropic_anl_pos}(b) ($L_{m}$ is constant up to 5 digits). In this figure, the attenuation length is slightly decreasing ($q_{m}$ is slightly increasing) for all the modes under investigation. Consequently, with the increase of the power, the field profiles will be more localized at the core metal interfaces with a nearly constant portion in the metal claddings (see figure~\ref{fig:Ex_field_profiles_anl_pos}). Therefore, the overall losses decrease with the increase of the power, as it can be seen in figure~\ref{fig:nonlinear_dispersion_curves_anisotropic_anl_pos}(c). It is worth mentioning that this interesting property cannot be obtained in the usual isotropic NPSWs, where the losses increase with the power~\cite{Walasik15b}.   
  In figure~\ref{fig:Ex_field_profiles_anl_pos}, we present the $E_{x}$ components for the symmetric (blue), asymmetric (green), and the antisymmetric (red) modes, obtained from  Full NL FEM. In one hand, the field profiles in the metal claddings are nearly constant as we discussed above, on the other hand, the intensities at the metal/core interfaces are increasing with the power. Consequently, the losses decrease with the increase of the power, as it is shown in figure~\ref{fig:nonlinear_dispersion_curves_anisotropic_anl_pos}(c). 
We focused only on the main modes, higher order modes exist and they are quite similar to the ones obtained in the isotropic case~\cite{Walasik15b}, but due to the enhancement of the nonlinearity, these nonlinear higher order modes appear at lower powers compared to the isotropic case. Their full investigation is beyond the scope of this study. 
\subsection{Beyond the EJEM assumptions}
\label{subsec:beyond_EJEM_assump}
\begin{figure}[h] 
	\centering
	\includegraphics[width=0.9\columnwidth,angle=0,clip=true,trim= 0 0 0 0 0]{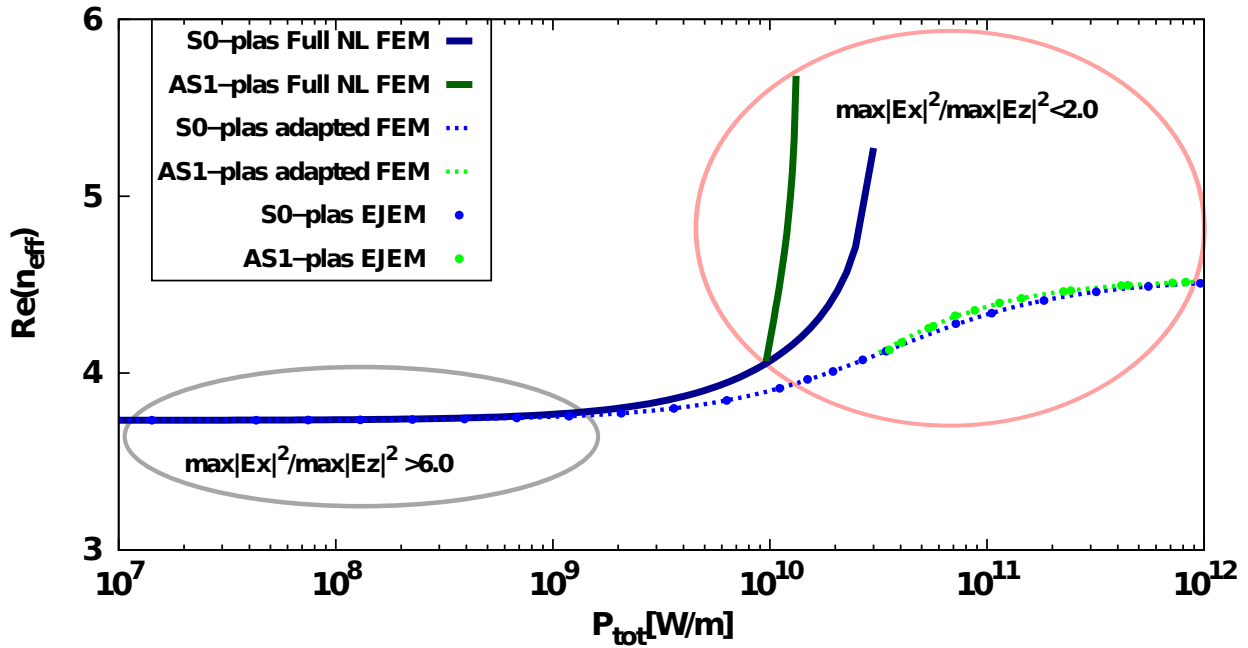}
	\caption{Nonlinear dispersion curves for the anisotropic case with $\epsilon_{xx}=3.46^{2}$, $\epsilon_{zz}=5.0$, $\alpha_{xx}=\alpha_{zz}=3.18 \times 10^{-19} m^{2}/V^{2}$. Solid lines for the results obtained from the Full NL FEM, closed circles for EJEM, and dashed lines for the adapted FEM. Blue, green colours for the nonlinear symmetric S0-plas, and the nonlinear asymmetric AS1-plas, respectively}
	\label{fig:limit_EJEM_epszz_GT_epsxx}
\end{figure} 
We have seen in subsections~\ref{subsubsec:valid_isotropic},~\ref{subsec:numerical_example_anl_neg}, and~\ref{subsec:numerical_example_anl_pos}, that the results obtained from the semi-analytical EJEM and the adapted FEM, qualitatively agree with the results obtained from the more accurate model Full NL FEM. More precisely, when the EJEM assumptions shown in subsection~\ref{subsec:EJEM} are valid, the three methods produce the same results. Nevertheless, when the EJEM assumptions start to be partially not valid (at high powers), we have found some discrepancies between the results obtained from the three methods (see figures~\ref{fig:nonlinear_dispersion_curves_isotropic},~\ref{fig:nl_dispersion_anisotropic_NEGATIVE_anl}, and~\ref{fig:nonlinear_dispersion_curves_anisotropic_anl_pos}). However, the global behaviour  of the nonlinear dispersion curves remains the same. Consequently, using the semi-analytical EJEM, we were able to get more insights into the computed nonlinear solutions. In this subsection, we present a numerical example for the anisotropic case in order to show the limitations of the semi-analytical EJEM. In figure~\ref{fig:limit_EJEM_epszz_GT_epsxx}, we choose $\epsilon_{xx}=3.46^{2}$, $\epsilon_{zz}=5.0$, $\alpha_{xx}=\alpha_{zz}=3.18 \times 10^{-19}$ $m^{2}/V^{2}$. In this case, the effective nonlinearity in the EJEM framework is positive $a^{\texttt{\tiny {EJEM}}}_{nl}>0$. We consider only the nonlinear symmetric (S0-plas) and asymmetric (AS1-plas) modes in this discussion. At low powers, when the transverse component of the electric field $E_{x}$ is much higher than the longitudinal one $E_{z}$, all the methods coincide. However, at high powers, we clearly see that the results obtained from the Full NL FEM (solid curves) differ quantitatively from the results obtained from the EJEM (closed circles) and from the adapted FEM (dashed lines), since at high powers, unlike figures~\ref{fig:nonlinear_dispersion_curves_isotropic},~\ref{fig:nl_dispersion_anisotropic_NEGATIVE_anl}, and~\ref{fig:nonlinear_dispersion_curves_anisotropic_anl_pos}, the ratio max$|E_{x}|^{2}/|E_{z}|^{2}$ is less than 2.0, this inequality means that the amplitude of $E_{z}$ is comparable to those of $E_{x}$ implying that the main assumption of the EJEM (that the Kerr-nonlinearity depends only on $E_{x}$) is no longer valid. While, in the Full NL FEM, we take into account both $E_{x}$ and $E_{z}$ in the Kerr-type nonlinearity. This example indicates that the classification of the nonlinear solutions according to the sign of $a^{\texttt{\tiny {EJEM}}}_{nl}$  given by equation~(\ref{eq:a_nl_ansio}) derived in the frame of the EJEM cannot be extended to all the possible solutions computed within the Full NL FEM. Nevertheless, this classification remains useful, since it covers a wide range of configurations, and being analytical, it is straightforward to define it. This remark illustrates the complementarity of the two methods we described and validated in this study.

\section{Conclusion}
We have presented two different models to study the stationary TM nonlinear solutions propagating in  slot plasmonic waveguides with a nonlinear anisotropic metamaterial core. The first model is a semi-analytical one in which we describe the nonlinear electromagnetic field components in the waveguide by the Jacobi elliptical functions, it extends to the anisotropic case our previous model valid only for isotropic configuration. Within this model, we have studied the nonlinear dispersion relations for all the possible feasible solutions propagating in such structures. The nonlinearity in this method is treated in an approximated manner in which only the transverse component is used in the Kerr-nonlinearity term. In addition, we must assume that the nonlinear refractive index change is small compared to the linear one. The second model is based on the finite-element method, in which all the electromagnetic field components are computed numerically. In this method, the nonlinearity is treated correctly, and all the electric field components are used in the Kerr-nonlinearity without any further assumptions. Finally, we presented several numerical results to validate our models and illustrate their limitations. We have found that the methods provide identical results at low powers, while at high power there exist some discrepancies between them due to the assumptions required by the semi-analytical model. This semi-analytical approach EJEM gives more insights into the behaviour  of the nonlinear dispersion curves and the field profiles than the more numerical FEM one, while the FEM describes the nonlinearity in proper way, the only disadvantage is that the field profiles are computed numerically. We have also demonstrated the influence of the linear and nonlinear anisotropic terms of the permittivity on the nonlinear dispersion curves and on the field profiles for two different cases defined in the frame of the first model. First, for $a^{\texttt{EJEM}}_{nl} <0$, we have found that the nonlinear metamaterial core with positive Kerr-type nonlinearity can act as a defocusing medium. Second, for $a^{\texttt{EJEM}}_{nl} >0$, we have presented one interesting example providing a huge enhancement of the nonlinearity, moreover, losses are decreasing with the power unlike the usual isotropic plasmonic slot waveguides.

\section*{Funding}
G. R. would like to thank the PhD school ED 352 "Physique et Sciences de Mati\`ere" and the International Relation Service of Aix-Marseille University (projectAAP 2015 "Noliplasmo 2D") for their respective fundings.


\appendix


\section{Unbounded solutions in the EJEM}
 \label{appendixI:unbounded_solutions}

In this appendix, we briefly describe the unbounded solutions obtained in the case $a^{\texttt{\tiny {EJEM}}}_{nl}<0$ presented in subsection~\ref{subsec:negative_anl}. We have mentioned before that the only bounded solution with finite electromagnetic energy in the nonlinear core for  $a^{\texttt{\tiny {EJEM}}}_{nl}<0$ can be obtained only when $q^2_{core}<0$ (see~\eqref{eq:q_sqr_different_layers}) and $C_{0}>0$ (see~\eqref{eq:C0_cont_conditions_negative_anl}) under the condition shown in~\eqref{eq:condition_real_Hy_caseIa}. 

In this appendix, we will study the other possible subcases in the frame of $a^{\texttt{\tiny {EJEM}}}_{nl}<0$. First, we begin with the subcase in which the condition shown in~\eqref{eq:condition_real_Hy_caseIa} is not satisfied. Since we are only looking for a real solution for~\eqref{eq:first_integral_negative_a_nl_negative_q_sqr_core_pos_C0}, we can write the nonlinear wave equation as:
\begin{align}\label{eq:first_integral_negative_a_nl_neg_condition_not_satisfied} 
\frac{d H_{y}}{\sqrt{\gamma^4 -\delta^2 H_{y}^{2}+H_{y}^4(x)}}=\pm \sqrt{\frac{1}{A}}dx,
\end{align}
where $\gamma^4=AC_{0}$ and $\delta^2=AQ$
The coefficients $\gamma^4 >0$ and $\delta^2>0$ since A, Q, and $C_{0}$ are positive quantities in this subcase (see subsection~\ref{subsec:negative_anl}). Integrating~\eqref{eq:first_integral_negative_a_nl_neg_condition_not_satisfied} and using the relation 263.00 from~\cite{Byrd54-Handbook-elliptic-integrals} yields the magnetic field profile in the nonlinear core
\begin{equation}
\label{eq:Hy_field_caseIa_condition_not_satisfied}
\begin{aligned}
H^2_{y}(x)=\gamma^2 \frac{1+\text{cn}\left\lbrace \right\rbrace}{1-\text{cn}\left\lbrace \right\rbrace}, ~~~~~~~~~~~~~~~~~~~~~~~~~~~~~~\\
\left. \text{cn}\left\lbrace \right\rbrace= \text{cn} \left[\mp 2 \gamma \sqrt{\frac{1}{A}} \left(x+\frac{d_{core}}{2} \right) +X_{0} \right\vert \text{m}    \right],
\end{aligned}
\end{equation}
where, 
\begin{align}
\label{eq:m_x0_caseIa_condition_not_satisfied}
~\text{m}=\frac{2\gamma^2+\delta}{4\gamma^2},~~~
X_{0}=\left. \text{cn}^{-1}\left[  \frac{H^2_{y}|_{(x=-{d_{core}}/{2})} - \gamma^2}{H^2_{y}|_{(x=-{d_{core}}/{2})} + \gamma^2} \right\vert \text{m} \right].
\end{align}
\begin{figure}[h]
	\centering
	\includegraphics[width=0.85\columnwidth,angle=0,clip=true,trim= 0 0 0 0]{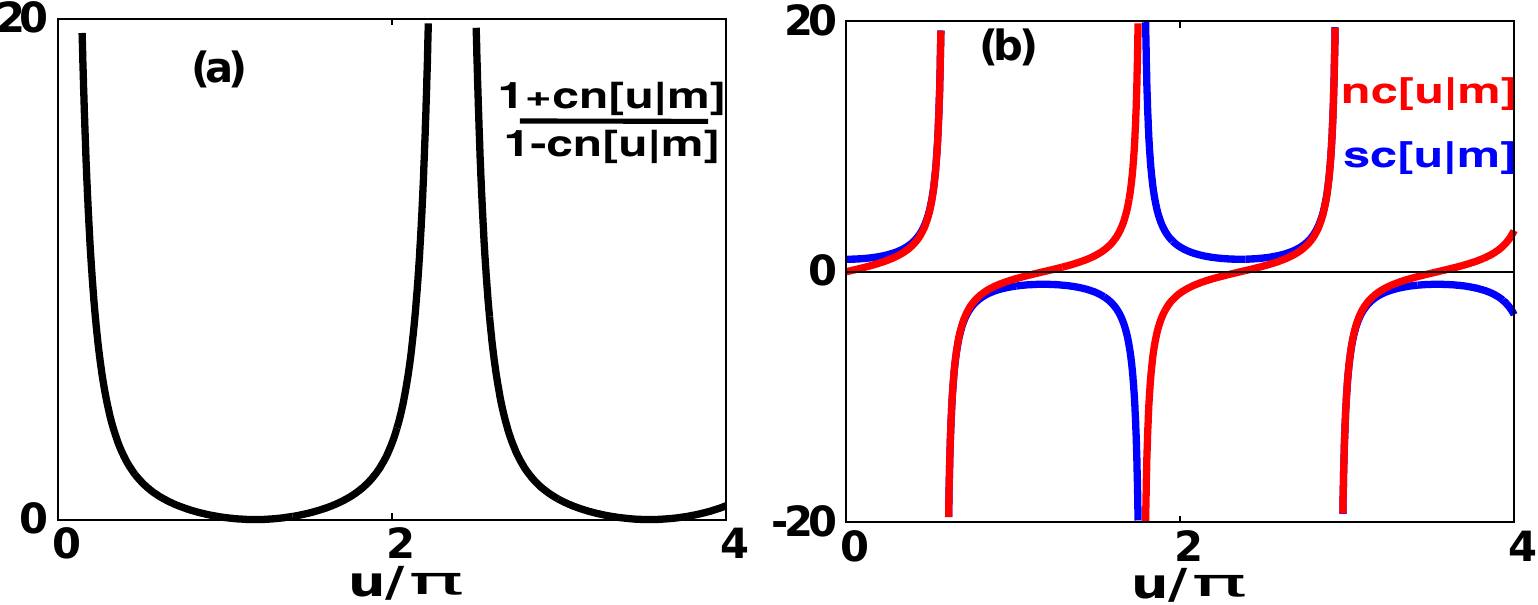}
	\caption{Unbounded periodic Jacobi elliptical functions with argument $u$ and parameter $m=0.5$. The global shape of the functions does not change $\forall m \in [0,1]$.}
	\label{fig:unbounded_Jacobi_elliptical_functions}
\end{figure}
\begin{table*}[t]
	\centering
	\caption{\bf The unbounded magnetic field profiles with their reduced parameters for the case $a^{\texttt{\tiny{EJEM}}}_{nl}<0$. The reduced parameter $A={2}/ \left( {k_{0}^2|a^{\texttt{\tiny {EJEM}}}_{nl}|} \right)$ is fixed for all the cases, while
		$Q=k_{0}^2 |q^{2}_{core}|$ for $q^2_{core}<0$ and $Q=k_{0}^2 q^{2}_{core}$ for $q^2_{core}>0$.}
	\begin{tabular}{p{0.05\textwidth}p{0.04\textwidth}p{0.17\textwidth}p{0.17\textwidth}p{0.06\textwidth}p{0.18\textwidth}c}
		\hline
		$q^2_{core}$ &~$C_{0}$ & ~~~~~~~~$\gamma^2$&~~~~~~~~~~ $\delta^2$ &~ m &~~~~~~~~~~~~~ $X_{0}$&$H_{y}(x)$ \\
		\hline 
		$<0$ & $<0$ &$\frac{-AQ+\sqrt{(AQ)^2+4A|C_{0}|}}{2}$ & $\frac{+AQ+\sqrt{(AQ)^2+4A|C_{0}|}}{2}$& $\frac{\gamma^2}{\gamma^2+\delta^2}$& $\left. \text{nc}^{-1} \left[ \frac{H_{[-{d_{core}}/{2}]}}{\delta} \right\vert \text{m} \right]$ &$\begin{small}
		\left. \delta \text{nc} \left[  \pm \sqrt{\frac{\gamma^2+\delta^2}{A}}\left(x+\frac{d_{core}}{2} \right)+X_{0} \right\vert \text{m}\right] \end{small}$ \\\\
		$>0$ & $>0$ &$\frac{AQ+\sqrt{(AQ)^2-4AC_{0}}}{2}$ &$\frac{AQ-\sqrt{(AQ)^2-4AC_{0}}}{2}$&$\frac{\gamma^2-\delta^2}{\gamma^2}$&$\left. \text{sc}^{-1}\left[  \frac{H_{[-{d_{core}}/{2}]}}{\delta}  \right\vert \text{m} \right]$& $\left. \delta \text{sc} \left[  \pm \sqrt{\frac{\gamma^2+\delta^2}{A}}\left(x+\frac{d_{core}}{2} \right)+X_{0} \right\vert \text{m}\right]$ \\\\
		$>0$ & $<0$ & $\frac{AQ+\sqrt{(AQ)^2+4A|C_{0}|}}{2}$&$ \frac{-AQ+\sqrt{(AQ)^2+4A|C_{0}}|}{2}$ &$ \frac{\gamma^2}{\gamma^2+\delta^2}$&$ \left. \text{nc}^{-1}\left[  \frac{H_{[-{d_{core}}/{2}]}}{\delta} \right\vert \text{m} \right]$&$\left. \delta \text{nc} \left[\pm \sqrt{\frac{\gamma^2+\delta^2}{A}}\left(x+\frac{d_{core}}{2} \right)+X_{0} \right\vert \text{m}    \right]$\\
		\hline
	\end{tabular}
	\label{tab:Hy_unbounded_negative_anl}
\end{table*}

The above solution represents the magnetic field component in the nonlinear core for $a^{\texttt{\tiny {EJEM}}}_{nl}<0$, $q^2_{core}<0$, and $C_{0}>0$  when the condition~\eqref{eq:condition_real_Hy_caseIa} is not satisfied which means that $q^{4}_{core} < {2 C_{0} |a^{\texttt{\tiny{EJEM}}}_{nl}| }/{k_{0}^2}$. This kind of solution is described by periodic and unbounded function (see figure~\ref{fig:unbounded_Jacobi_elliptical_functions}(a)) such that the electromagnetic energy in the core is infinite. A similar unbounded solution has already been obtained in nonlinear dielectric waveguides~\cite{Maradudin88}.

Next, we consider the other subcases in which we get unbounded solutions for $a^{\texttt{\tiny {EJEM}}}_{nl}<0$. Table~\ref{tab:Hy_unbounded_negative_anl} shows their unbounded magnetic field profiles with their reduced parameters according to the signs of $q^2_{core}$ and $C_{0}$. The first two columns of Table~\ref{tab:Hy_unbounded_negative_anl}  give the signs of $q^2_{core}$ and $C_{0}$, and the last column represents the corresponding magnetic field profiles in terms of one of the unbounded Jacobi elliptical functions nc$[u|m]$ or sc$[u|m]$~\cite{David-mathematica07,Salas14-duff-eq} with argument $u$ and parameter $m$ (see also figure~\ref{fig:unbounded_Jacobi_elliptical_functions}(b)). The other columns correspond to the reduced parameters used to describe the magnetic field profiles shown in the last column. We have used the same procedure to derive the nonlinear magnetic field profiles as in subsections~\ref{subsec:negative_anl},~\ref{subsec:positive_anl} and references~\cite{Walasik15a,Walasik15b}. One can see that the first and the third subcases can be described by the same unbounded function nc[u$|$m] however, the reduced parameters $\gamma^2$ and $\delta^2$ are reversed. In our study, we will exclude the solutions depicted in Table~\ref{tab:Hy_unbounded_negative_anl} since we are looking for a guided nonlinear wave propagating with finite energy in the core.


\section{Adapted FEM to match EJEM assumptions} 
\label{appendixII:adapted_FEM}
In this appendix, we show that the FEM presented in subsection~\ref{subsec:FEM} can be compared to the semi-analytical approach described in subsection~\ref{subsec:EJEM} by taking into account all the EJEM assumptions. In the frame of the EJEM, the nonlinearity depends only on the transverse component of the electric field $E_{x}$ and the nonlinear refractive index change is small compared to the linear one. Therefore, one way to tackle this problem is  to consider only the weak formulation for full (TM) wave equation in terms of the magnetic field forcing both the inhomogeneous permittivity term induced by the nonlinearity and the structure interfaces to take into account the EJEM assumptions
\begin{equation}
\label{eq:weak_formualtion_FEM_adapted_EJEM}
\begin{aligned}
\frac{-1}{k^2_{0}}\int_{\Gamma}  \frac{1}{\epsilon_{zz}(x)} \frac{d h_{y}}{d x} \frac{d h^{\prime}_{y}}{d x} dx +\int_{\Gamma} h_{y}(x) h^\prime_{y}(x) dx~~~~~~~~~~~~~~~~~~~~~~~~~~~~~~~~~~~~~~~~~~~~~~\\       
- \int_{\Gamma}\frac{ |E_{x}|^{2} }{\epsilon_{xx}^{2} \epsilon_{zz}} \left( n_{eff}^{2} \left( \alpha_{zz}\epsilon_{xx}-\alpha_{xx}\epsilon_{zz}\right)-\alpha_{zz}\epsilon_{xx}^{2}   \right) h_{y}(x) h^{\prime}_{y}(x) dx~~~~~~~~~~~~~ \\
=n_{eff}^2 \int_{\Gamma} \frac{1}{\epsilon_{xx}(x)} h_{y}(x) h^\prime_{y}(x)dx,~~~~~~~~~~~~~~~~~    
\end{aligned}
\end{equation}
such that, $\forall$ $h^\prime_{y}(x)\in \mathtt{H}_{0}^{1}(\gamma)$ we look for $h_{y}(x) \in \mathtt{H}_{0}^{1}(\gamma)$, where  $\mathtt{H}_{0}^{1}(\gamma)$ is the Sobolev space of order $1$ with null Dirichlet boundary conditions on the domain of integration $\Gamma$. The nonlinear parameters $\alpha_{xx}$ and $\alpha_{zz}$ are non-zero only in the core. In the frame of our EJEM assumptions, we assumed that the nonlinear refractive index change is small compared to the linear one, and $E_{x}$ and $E_{z}$ depend only on the linear part of the permittivity $\epsilon_{xx}$ and $\epsilon_{zz}$, respectively. This kind of approximation was already used in~\cite{Stegeman85,Ariyasu85}. The algorithm used to solve~\eqref{eq:weak_formualtion_FEM_adapted_EJEM} is quite similar to algorithm~\ref{alg: fixed_power_full_nl} except that we directly plug the initial field $E_{x}$ in the second term of~\eqref{eq:weak_formualtion_FEM_adapted_EJEM} and the rescaling factor $\chi$ is computed at each iteration for a given fixed power $P_{tot}$ from the following approximated relation:
\begin{equation}
\label{eq:chi_Ex_lin_epsxx}
P_{tot}= \frac{\Re e(n_{eff}) \chi^2}{2\epsilon_{0}\epsilon_{xx}c}  \int_{\Gamma} |h_{y}(x)|^{2} dx,   
\end{equation}
where $h_{y}$ and $n_{eff}$ are the outputs at each iteration. 
\\
Finally, we would like to mention that the single-component eigenvalue problem shown in~\eqref{eq:weak_formualtion_FEM_adapted_EJEM} or in~\cite{Walasik14}, in the frame of our FEM formalism using the fixed power algorithm is not possible to be solved taking into account all the electric field components in the Kerr-type nonlinearity. To clarify this point, we begin with the power density transmitted per unit length along $y$ direction (the longitudinal $z$ component of the pointing vector integrated over the transverse $x$-direction)
	\begin{equation} \label{eqn:chi_Ex_full_single_wave_eq}
	P_{tot}=\frac{\Re e(n_{eff})}{2\epsilon_{0}c} \int_{\Gamma} \frac{\chi^2}{\epsilon_{xx}+\alpha_{xx} \left(|E_{x}|^{2} + |E_{z}|^{2} \right)} |h_{y}(x)|^{2} dx,
	\end{equation} 
 where $h_{y}$ is the eigenvector of the single-component eigenvalue problem. In order to obtain the rescaling factor $\chi$, we need to express $E_{x}$, $E_{z}$ in terms of the eigenvector $h_{y}$ and $\chi$, which is not possible since both $E_{x}$ and $E_{z}$ depend on the nonlinear permittivity $\epsilon_{x}$ and $\epsilon_{z}$, respectively, as it is shown in equations~(\ref{eq:maxwell_TM}). Another way to solve this problem is to write the weak formulation in terms of the electric field components $E_{x}$ and $E_{z}$, and it then possible to obtain an explicit formula for $\chi$ as a function of $P_{tot}$. But, this approach requires the use of discontinuous Galerkin method in the FEM~\cite{lanteri08_time_Harmonic_DG,lanteri13_2D_DG_time_Harmonic} since $E_{x}$ is a discontinuous function, which is beyond the scope of the present study. Consequently, in order to take into account all the electric field components in the Kerr-nonlinearity in our FEM using the fixed power algorithm, we must solve a coupled eigenvalue problem in terms of the continuous electromagnetic field components $E_{z}$ and $H_{y}$ as described in subsection~\ref{subsec:FEM}.

\bibliographystyle{unsrt}

\end{document}